\documentclass[journal,twoside,twocolumn,letterpaper]{IEEEtran}
\usepackage{amsmath,amssymb,amscd,latexsym,dsfont}
\usepackage{multirow}
\usepackage{comment}
\usepackage{color}
\usepackage{graphicx}
\usepackage{subfigure}
\usepackage{enumerate}
\usepackage{stfloats}
\usepackage{psfrag}
\usepackage{setspace}
\usepackage{cite}
\usepackage{balance}

\usepackage{tikz,pgfplots}
\usetikzlibrary{positioning,shadows,shapes,fit,arrows,backgrounds,calc}
\colorlet{figyellow}{yellow!40!white}
\colorlet{figred}{red!40!white}
\colorlet{figblue}{blue!20!white}
\colorlet{figgreen}{green!30!white}
\colorlet{figgray}{black!10!white}
\colorlet{lightgray}{gray!40!white}
\colorlet{lightblue}{blue!40!white}
\colorlet{lightred}{red!40!white}
\colorlet{lightgreen}{green!40!white}

\usepgfplotslibrary{external} 
\usepackage{algorithmic}
\usepackage[]{algorithm}
\IEEEoverridecommandlockouts

\newcommand{\MED}{d}

\newcommand{\mcIX}{\mc{I}}

\newcommand{\pdf}{f}

\newcommand{\Pxi}{p_i}

\newcommand{\Pxj}{p_j}

\newcommand{\sumi}{\sum_{i\in\mcIX}}

\newcommand{\sumjni}{\sum_{\substack{j\in\mcIX\\j\neq i}}}

\newcommand{\sumjnid}{\sum_{\substack{j\in\mcIX\\ \delta_{ij}=\MED}}}
\newcommand{\sumijnid}{\sum_{\substack{i,j\in\mcIX\\ \delta_{ij}=\MED}}}

\newcommand{\sumijd}{\sum_{\substack{i,j\in\mcIX\\ \delta_{ij}=\MED}}}
\newcommand{\sumjid}{\sum_{\substack{j,i\in\mcIX\\ \delta_{ji}=\MED}}}

\newcommand{\deltaij}{\delta_{ij}}
\newcommand{\bdij}{\bd_{ij}}

\newcommand{\bdin}{\bd_{in}}

\newcommand{\Cij}{\mcC_{ij}}
\newcommand{\Cin}{\mcC_{in}}

\newcommand{\hd}[2]{\mathop{\gamma_{#1#2}}\nolimits}	

\newcommand{\tp}[2]{\mathop{T_{#1#2}}\nolimits} 
\newcommand{\tpMAP}[2]{\mathop{T_{#1#2}^{\tnr{map}}}\nolimits} 
\newcommand{\tpML}[2]{\mathop{T_{#1#2}^{\tnr{ml}}}\nolimits} 

\newcommand{\limsnszero}[1]{{\lim_{\SZ\rightarrow0}{#1}}}

\newcommand{\tr}[1]{\mathrm{#1}}

\newcommand{\mc}[1]{\mathcal{#1}}

\newcommand{\set}[1]{\{#1\}}
\newcommand{\Set}[1]{\bigl\{#1\bigr\}}

\newcommand{\ld}{\ldots}

\newcommand{\ov}[1]{\overline{#1}}

\newcommand{\ie}{i.e.,~}

\newcounter{lemma}
\newtheorem{theorem}[lemma]{Theorem}
\newtheorem{corollary}[lemma]{Corollary}
\newtheorem{lemma}{Lemma}

\newtheorem{exampleplain}{Example}

\newenvironment{example}{\begin{exampleplain}}{~\hfill$\vartriangle$\end{exampleplain}}

\newcommand{\tnr}[1]{\textnormal{#1}}
\newcommand{\inner}[2]{\left\langle#1,#2\right\rangle}

\newcommand{\BEP}[0]{P_{\tnr{b}}}

\newcommand{\SZ}[0]{\sigma}
\newcommand{\SX}[0]{E_{\tnr{s}}}

\newcommand{\hbX}{\hat{\bX}}

\newcommand{\hbXMAP}{\hat{\bX}^{\tnr{map}}}

\newcommand{\hbXML}{\hat{\bX}{}^{\tnr{ml}}}
\newcommand{\BEPxpMAP}{\BEP(\SZ)}
\newcommand{\BEPijMAP}[0]{\tp{i}{j}(\SZ)}

\newcommand{\QF}[0]{{Q}}

\newcommand{\bX}{\boldsymbol{X}}

\newcommand{\bZ}{\boldsymbol{Z}}

\newcommand{\bY}{\boldsymbol{Y}}

\newcommand{\bd}{\boldsymbol{d}}

\newcommand{\mcC}{\mc{C}}

\newcommand{\mcH}{\mc{H}}
\newcommand{\mcI}{\mc{I}}

\newcommand{\mcR}{\mc{R}}

\newcommand{\mcX}{\mc{X}}

\newcommand{\bC}{\boldsymbol{C}}

\newcommand{\bx}{\boldsymbol{x}}
\newcommand{\bc}{\boldsymbol{c}}

\newcommand{\by}{\boldsymbol{y}}

\newcommand{\argmax}{\mathop{\mathrm{argmax}}}

\newcommand{\explow}[1]{\mathop{\mathrm{exp}\!}\left({#1}\right)}
\newcommand{\figref}[1]{Fig.~\ref{#1}}

\newcommand{\secref}[1]{Section~\ref{#1}}
\newcommand{\theoref}[1]{Theorem~\ref{#1}}

\newcommand{\lemmaref}[1]{Lemma~\ref{#1}}
\newcommand{\lemmasref}[2]{Lemmas~\ref{#1} and \ref{#2}}
\newcommand{\cororef}[1]{Corollary~\ref{#1}}

\title{Asymptotic Comparison of ML and MAP Detectors for Multidimensional Constellations}
\author{
Alex Alvarado,~\IEEEmembership{Senior Member,~IEEE}, 
Erik Agrell,~\IEEEmembership{Senior Member,~IEEE},
and Fredrik Br\"annstr\"om,~\IEEEmembership{Member,~IEEE},
\thanks{Research supported in part by the Swedish Research Council (under grant \#2011-5950), by the Ericsson Research Foundation, and by the Engineering and Physical Sciences Research Council (EPSRC) project UNLOC (EP/J017582/1).}
\thanks{A.~Alvarado was with the Optical Networks Group, Department of Electronic \& Electrical Engineering, University College London, London WC1E 7JE, UK. He is now with the Signal Processing Systems Group, Department of Electrical Engineering, Eindhoven University of Technology (TU/e), Eindhoven, 5600 MB, The Netherlands (email: alex.alvarado@ieee.org).}
\thanks{E.~Agrell and F.~Br\"annstr\"om are with the Department of Electrical Engineering, Chalmers University of Technology, SE-41296 G\"oteborg, Sweden (emails: \set{agrell,\,fredrik.brannstrom}@chalmers.se).}
}
\begin{document}
\maketitle

\begin{abstract}
A classical problem in digital communications is to evaluate the symbol error probability (SEP) and bit error probability (BEP) of a multidimensional constellation over an additive white Gaussian noise channel. In this paper, we revisit this problem for nonequally likely symbols and study the behavior of the optimal maximum a posteriori (MAP) detector at asymptotically high signal-to-noise ratios. Exact closed-form asymptotic expressions for SEP and BEP for arbitrary constellations and input distributions are presented. The well-known union bound is proven to be asymptotically tight under general conditions. The performance of the practically relevant maximum likelihood (ML) detector is also analyzed. Although the decision regions with MAP detection converge to the ML regions at high signal-to-noise ratios, the ratio between the MAP and ML detector in terms of both SEP and BEP approach a constant, which depends on the constellation and a priori probabilities. Necessary and sufficient conditions for asymptotic equivalence between the MAP and ML detectors are also presented.
\end{abstract}


\section{Introduction}\label{Sec:Introduction}

The evaluation of the symbol error probability (SEP) and bit error probability (BEP) of a multidimensional constellation over an additive white Gaussian noise (AWGN) channel is a classical problem in digital communications. This problem traces back to \cite{Gilbert52} in 1952, where upper and lower bounds on the SEP of multidimensional constellations based on the maximum likelihood (ML) detector were first presented. 

When nonuniform signaling is used, \ie when constellation points are transmitted using different probabilities, the optimal detection strategy is the maximum a posteriori (MAP) detector. The main drawback of MAP detection is that its implementation requires decision regions that vary as a function of the signal-to-noise ratio (SNR). Practical implementations therefore favor the (suboptimal) ML approach where the a priori probabilities are essentially ignored. For ML detection, the decision regions are the so-called Voronoi regions, which do not depend on the SNR, and thus, are simpler to implement.

Error probability analysis of constellations for the AWGN channel has been extensively investigated in the literature, see e.g., \cite{Wozencraft65_Book,Jacobs67,Foschini74,Smith75,Forney89a,Kschischang93}. In fact, this a problem treated in many---if not all---digital communication textbooks. To the best of our knowledge, and to our surprise, the general problem of error probability analysis for multidimensional constellations with arbitrary input distributions and MAP detection has not been investigated in such a general setup.

As the SNR increases, the MAP decision regions tend towards the ML regions. Intuitively, one would then expect that both detectors are asymptotically equivalent, which would justify the use of ML detection. In this paper, we show that this is not the case. MAP and ML detection give different SEPs and BEPs asymptotically, where the difference lies in the factors before the dominant Q-function expression. More precisely, the ratio between the SEPs with MAP and ML detection approaches a constant, and the ratio between their BEPs approaches another constant. These constants are analytically calculated for arbitrary constellations, labelings, and input distributions. To the best of our knowledge, this has never been previously reported in the literature. Numerical results support our analytical results and clearly show the asymptotic suboptimality of ML detection.

All the results in this paper are presented in the context of detection of constellations with arbitrary number of dimensions. These multidimensional constellations, however, can be interpreted as finite-length codewords from a code, where the cardinality of the constellation and its dimensionality correspond to the codebook size and codeword length, respectively. In this context, the results of this paper can be used to study the performance of the MAP and ML \emph{sequence decoders} at asymptotically high SNRs. 

This paper is organized as follows. In \secref{Sec:Preliminaries}, the model is introduced and in \secref{Sec:Bounds}, the error probability bounds are presented. The main results of this paper are given in \secref{Sec:Asymptotics}. Conclusions are drawn in \secref{Sec:Conclusions}. All proofs are deferred to Appendices.

\section{Preliminaries}\label{Sec:Preliminaries}

\subsection{System Model}\label{Sec:Preliminaries:Model}

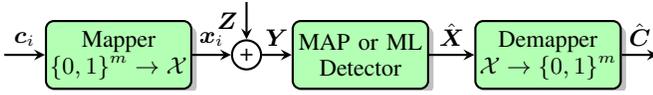
\begin{figure}[tpb]
\newlength{\blocksep}\setlength{\blocksep}{6mm} 
\begin{center}
\small{
\hspace{-2ex}
\begin{tikzpicture}[>=stealth,auto,tight background,
block/.style={rectangle,rounded corners=3pt,thick,draw,inner sep=1.5pt,minimum width=18mm,
minimum height=9mm,fill=figyellow,drop shadow,align=center,execute at begin node=\setlength{\baselineskip}{2.5ex}},
plain/.style={align=center,execute at begin node=\setlength{\baselineskip}{5.5ex}}
]
\node[plain] (C) {}; 
\coordinate (AWGN) at ($(C)+(-0.35cm,0)$);
\node[draw,thick,circle,fill=white,inner sep=1.5pt] (N1) at (AWGN) {+};
\node[plain,above=20pt of AWGN] (Z) {};
\node[block,fill=figgreen,left=1.2\blocksep of AWGN] (M) {Mapper\\ $\set{0,1}^{m} \rightarrow \mcX$}; 
\node[block,fill=figgreen,right=\blocksep of AWGN] (HD) {MAP or ML\\Detector}; 
\node[plain,left=0.9\blocksep of M] (SRC) {};
\node[block,fill=figgreen,right=\blocksep of HD] (D) {Demapper\\ $\mcX \rightarrow \set{0,1}^{m}$}; 
\node[plain,right=0.8\blocksep of D] (SINK) {};
\draw[thick,->] (SRC) -- node[plain,above] {$\bc_{i}$} (M);
\draw[thick,->] (M) -- node[plain,above] {$\bx_{i}$} (N1);
\draw[thick,->] (N1) -- node[plain,above] {$\bY$} (HD);
\draw[thick,->] (Z) -- node[plain,left] {$\bZ$} (N1);
\draw[thick,->] (HD) -- node[plain,above] {$\hat{\bX}$} (D);
\draw[thick,->] (D) -- node[plain,above] {$\hat{\bC}$} (SINK);
\end{tikzpicture}
}
\end{center}
\caption{System model under consideration. For a given length-$m$ transmitted binary label $\bc_{i}$, the received vector $\bY$ is processed by the MAP or ML detector. The estimated symbol $\hat{\bX}$ is then converted to an estimated binary label $\hat{\bC}$.}
\label{model}
\end{figure}

The system model under consideration is shown in Fig.~\ref{model}.
We consider the discrete-time, real-valued, $N$-dimensional, AWGN channel
\begin{align}\label{AWGN}
\bY = \bX+\bZ,
\end{align}
where the transmitted symbol $\bX$ belongs to a discrete constellation $\mcX=\set{\bx_1,\bx_2,\ld,\bx_{M}}$ and $\bZ$ is an $N$-dimensional vector, independent of $\bX$, whose components are independent and identically distributed Gaussian random variables with zero mean and variance $\SZ^{2}$ per dimension. The conditional channel transition probability is
\begin{align}\label{pdf.channel}
\pdf(\by|\bx) = \frac{1}{(2\pi\SZ^{2})^{N/2}}\tr{exp}{\left(-\frac{\|\by-\bx\|^{2}}{2\SZ^{2}}\right)}.
\end{align}

We assume that the symbols are distinct and that each of them is transmitted with probability $\Pxi= \Pr\set{\bX=\bx_i}$, $0< \Pxi <1$. Neither the constellation points nor their probabilities depend on $\SZ$. We use the set $\mcIX= \set{1,\ld,M}$ to enumerate the constellation points. The average symbol energy is $\SX= \sumi \Pxi \|\bx_i\|^2 < \infty$. The Euclidean distance between $\bx_{i}$ and $\bx_{j}$ is defined as $\deltaij= \|\bx_i-\bx_j\|$ and the minimum Euclidean distance (MED) of the constellation as $\MED= \min_{i,j\in\mcIX:i\neq j}\deltaij$.

For the BEP analysis, assuming that $M$ is a power of two, we consider a binary source that produces length-$m$ binary labels. These labels are mapped to symbols in $\mcX$ using a \emph{binary labeling}, which is a one-to-one mapping between the $M=2^{m}$ different length-$m$ binary labels and the constellation points. The length-$m$ binary labels have an arbitrary input distribution, and thus, the same distribution is induced on the constellation points. The binary label of $\bx_{i}$ is denoted by $\bc_{i}$, where $i\in\mcIX$. The Hamming distance between $\bc_{i}$ and $\bc_{j}$ is denoted by $\hd{i}{j}$.

At the receiver, we assume that (hard-decision) symbol-wise decisions are made. The estimated symbol is then mapped to a binary label to obtain an estimate on the transmitted bits.\footnote{This detector based on symbols has been shown in \cite{Ivanov13a} to be suboptimal in terms of BEP; however, differences are expected only at high BEP values.}
For any received symbol $\by$, the MAP decision rule is\footnote{Throughout the paper, the superscripts ${\tnr{``map''}}$ and ${\tnr{``ml''}}$ denote quantities associated with MAP and ML detection, respectively.}
\begin{align}
\label{map}
\hbXMAP(\by)  	& = \argmax_{j\in\mcI}\set{\Pxj\pdf(\by|\bx_{j})}.
\end{align}
This decision rule generates MAP decision regions defined as
\begin{align}\label{map.j.region}
\mcR_{j}^{\tnr{map}}(\SZ) &= \set{\by\in\mathbb{R}^{N}: \Pxj\pdf(\by|\bx_{j}) \geq \Pxi\pdf(\by|\bx_{i}), \forall i\in\mcIX}
\end{align}
for all $j\in\mcIX$. Similarly, the ML detection rule is
\begin{align}
\label{ml}
\hbXML(\by)	&= \argmax_{j\in\mcI}\set{\pdf(\by|\bx_{j})},
\end{align}
which results in the decision regions
\begin{align}\label{ml.j.region}
\mcR_{j}^{\tnr{ml}}(\SZ) &= \set{\by\in\mathbb{R}^{N}: \pdf(\by|\bx_{j}) \geq \pdf(\by|\bx_{i}), \forall i\in\mcIX}.
\end{align}

\begin{figure}[tbp]
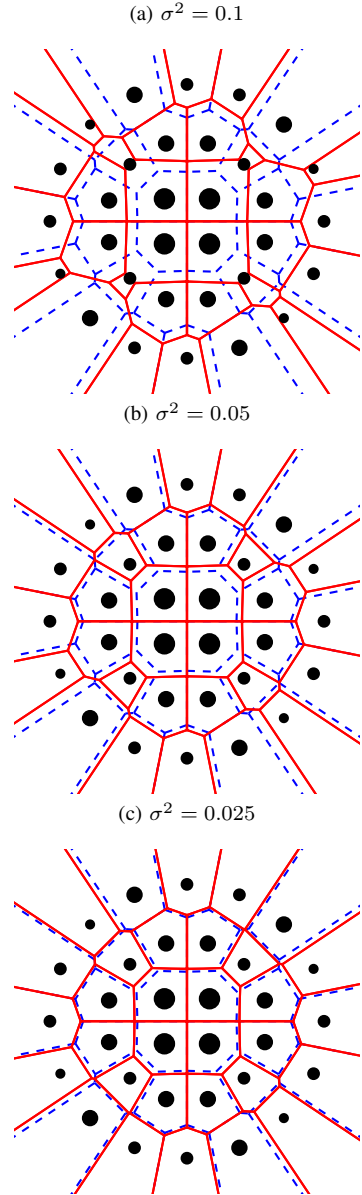

	\centering
	\begin{tikzpicture}
	\begin{axis}[axis lines=none,
	xminorgrids=true,
	width=0.34\textwidth,
	height=0.34\textwidth,
	xmin=-1.8,xmax=1.8,
	ymin=-1.8,ymax=1.8,
	xlabel style={yshift=0.1cm},
	xtick={-1.5,-0.5,0.5,1.5},
	every axis/.append style={font=\footnotesize},
	legend style={legend pos=south west,font=\scriptsize,legend cell align=left},
	title={(a) $\SZ^2=0.1$}
	] 
	\input{data/Valenti_32APSK_constellation.tikz}
	\input{data/Valenti_32APSK_constellation_Voronoi.tikz}
	\input{data/Valenti_32APSK_constellation_MAP_10_dB.tikz}
	\end{axis}
	\end{tikzpicture}
	\hspace{5pt}
	\begin{tikzpicture}
	\begin{axis}[axis lines=none,
	xminorgrids=true,
	width=0.34\textwidth,
	height=0.34\textwidth,
	xmin=-1.8,xmax=1.8,
	ymin=-1.8,ymax=1.8,
	xlabel style={yshift=0.1cm},
	xtick={-1.5,-0.5,0.5,1.5},
	every axis/.append style={font=\footnotesize},
	legend style={legend pos=south west,font=\scriptsize,legend cell align=left},
	title={(b) $\SZ^2=0.05$}
	]
	\input{data/Valenti_32APSK_constellation.tikz}
	\input{data/Valenti_32APSK_constellation_Voronoi.tikz}
	\input{data/Valenti_32APSK_constellation_MAP_13_dB.tikz}
	\end{axis}
	\end{tikzpicture}
	\hspace{5pt}
	\begin{tikzpicture}
	\begin{axis}[axis lines=none,
	xminorgrids=true,
	width=0.34\textwidth,
	height=0.34\textwidth,
	xmin=-1.8,xmax=1.8,
	ymin=-1.8,ymax=1.8,
	xlabel style={yshift=0.1cm},
	xtick={-1.5,-0.5,0.5,1.5},
	every axis/.append style={font=\footnotesize},
	legend style={legend pos=south west,font=\scriptsize,legend cell align=left},
	title={(c) $\SZ^2=0.025$}
	]
	\input{data/Valenti_32APSK_constellation.tikz}
	\input{data/Valenti_32APSK_constellation_Voronoi.tikz}
	\input{data/Valenti_32APSK_constellation_MAP_16_dB.tikz}
	\end{axis}
	\end{tikzpicture}
	\caption{ML (dashed blue) and MAP (solid red) decision regions for the constellation in Example~\ref{Example.Regions} and three values of the noise variance. The area of the constellation points is proportional to their probabilities. As the noise variance decreases, the MAP regions converge to the ML regions.}
	\label{Valenti_32APSK}
\end{figure}

\begin{example}\label{Example.Regions}
Consider the $32$-ary constellation with the nonuniform input distribution in \cite[Fig.~2, Table~I]{Valenti12}\footnote{Using three shaping bits, which results in radii $1,2.53, 4.30$.}. The constellation is shown in \figref{Valenti_32APSK}, where the area of the constellation points is proportional to the corresponding probabilities. In \figref{Valenti_32APSK}, the MAP and ML decision regions in \eqref{map.j.region} and \eqref{ml.j.region} are shown for three values of the noise variance. These results show how the MAP regions converge to the ML regions as the noise variance decreases.
\end{example}

\subsection{Error Probability}\label{Sec:Preliminaries:EP}

Throughout this paper, the SEP and BEP are denoted by $P_{\tnr{s}}(\SZ)$ and $P_{\tnr{b}}(\SZ)$, respectively. Furthermore, we are interested in the error probability (SEP and BEP) of both the MAP and ML detectors. To study these four error probabilities, we define the generic error probability function
\begin{align}\label{P}
P(\SZ) 	& = \sumi \Pxi \sumjni h_{ij} \tp{i}{j}(\SZ),
\end{align}
where the transition probability $\tp{i}{j}(\SZ)$ is given by
\begin{align}\label{Fij}
\tp{i}{j}(\SZ) 	& = \Pr\Set{\hbX(\bY)= \bx_{j}|\bX=\bx_{i}}\\
\label{Fij.2}
			& = \Pr\Set{\bY \in \mcR_{j}(\SZ)|\bX=\bx_{i}}.
\end{align}
The expressions \eqref{P}--\eqref{Fij.2} represent both the MAP and ML detectors, as well as both the SEP and BEP, as explained in the following.

The error probability with MAP detection is obtaining by using $\mcR_{j}(\SZ)=\mcR_{j}^{\tnr{map}}(\SZ)$ in \eqref{Fij.2}, where $\mcR_{j}^{\tnr{map}}(\SZ)$ is given by \eqref{map.j.region}. Similarly, the use of $\mcR_{j}(\SZ)=\mcR_{j}^{\tnr{ml}}(\SZ)$ in \eqref{Fij.2}, where $\mcR_{j}^{\tnr{ml}}(\SZ)$ is given by \eqref{ml.j.region}, leads to the error probability with ML detection.

To study the SEP, $h_{ij}$ in \eqref{P} should be set to one, which gives the well-known expression
\begin{align}\label{sep}
P_{\tnr{s}}(\SZ) 	= \sumi \Pxi \sumjni \tp{i}{j}(\SZ).
\end{align}
Similarly, the BEP expression \cite[Eq.~(1)]{Lassing03}, \cite[Eq.~(1)]{Agrell04}
\begin{align}\label{bep}
\BEPxpMAP 	& = \sumi \Pxi \sumjni \frac{\hd{i}{j}}{m} \BEPijMAP
\end{align}
is obtained by using $h_{ij}=\hd{i}{j}/m$ in \eqref{P}.
The four cases discussed above are summarized in the first three columns of Table~\ref{Table:EP}.

\begin{table}[tpb]
 \renewcommand\arraystretch{1.45}
	\caption{Values of $\hbX$ and $h_{ij}$ that used in \eqref{P}--\eqref{Fij.2} give SEP and BEP expressions for both the MAP and ML detectors. The last column shows the values of $w_{ij}$ for the asymptotic expressions in \secref{Sec:Asymptotics}.}
	\centering
	\begin{tabular}{ l l c c}
\hline

\hline
$P(\SZ)$					& $\hbX$ 		& $h_{ij}$						& $w_{ij}$  \\
\hline 

\hline
$P_{\tnr{s}}^{\tnr{map}}(\SZ)$ 	& $\hbXMAP$ 	& $1$ 						& $\sqrt{\dfrac{\Pxj}{\Pxi}}$\\ 
$P_{\tnr{s}}^{\tnr{ml}}(\SZ)$ 	& $\hbXML$ 	& $1$ 						& $1$\\ 
$P_{\tnr{b}}^{\tnr{map}}(\SZ)$ 	& $\hbXMAP$ 	& $\dfrac{\hd{i}{j}}{m}$ 	& $\sqrt{\dfrac{\Pxj}{\Pxi}}$\\ 
$P_{\tnr{b}}^{\tnr{ml}}(\SZ)$ 	& $\hbXML$ 	& $\dfrac{\hd{i}{j}}{m}$ 	& $1$\\ 
\hline
	\end{tabular}
	\label{Table:EP}
\end{table}

\section{Error Probability Bounds}\label{Sec:Bounds}

Error probability calculations for arbitrary multidimensional constellations and finite SNR are difficult because the decision regions defining the transition probabilities $\tp{i}{j}(\SZ)$ in \eqref{Fij.2} are in general irregular. Therefore, to analytically study the error probability, bounding techniques are usually the preferred alternative. In this section, we present two lemmas that give upper and lower bounds on the transition probability $\tp{i}{j}(\SZ)$. These bounds are expressed in terms of the Gaussian Q-function $\QF(x) = (1/\sqrt{2\pi}) \int_{x}^{\infty} \explow{-\xi^2/2}\tr{d}\xi$ and will then be used to upper- and lower-bound the SEP and BEP in \secref{Sec:Asymptotics}.

\begin{lemma}\label{Lemma.UB}
For any $i,j\in\mcIX$, $j\neq i$,
\begin{align}\label{Lemma.UB.eq}
\tp{i}{j}(\SZ) \leq \QF\biggl(\frac{\Delta_{ij}(\SZ)}{\SZ}\biggr),
\end{align}
where
\begin{align}\label{delta.ij}
\Delta_{ij}(\SZ)=
\begin{cases}
\frac{\deltaij}{2} \left(1+\frac{2\SZ^{2}}{\deltaij^{2}}\log\frac{\Pxi}{\Pxj}\right),	& \tnr{for MAP},\\
\frac{\deltaij}{2},	& \tnr{for ML}.
\end{cases}
\end{align}
\end{lemma}
\begin{IEEEproof}
See Appendix~\ref{Appendix.Lemma.UB}.
\end{IEEEproof}

\begin{lemma}\label{Lemma.LB}
For any $i,j\in\mcIX$, $j\neq i$ and any $\SZ<\tau_{ij}$,
\begin{align}\label{Lemma.LB.eq}
\tp{i}{j}(\SZ) \geq 
\begin{cases}
0, & \textnormal{if $\deltaij>\MED$},\\
\left(\QF\Bigl(\frac{\Delta_{ij}(\SZ)}{\SZ}\Bigr)-\QF\Bigl(\frac{\MED}{2\SZ}+\frac{r(\SZ)}{\sqrt{N}\SZ}\Bigr)\right)\cdot & \\
\qquad\qquad\quad \left(1-2\QF\left(\frac{r(\SZ)}{\sqrt{N}\SZ}\right)\right)^{N-1}, &\textnormal{if $\deltaij=\MED$},\\
\end{cases}
\end{align}
where $\Delta_{ij}(\SZ)$ is given by \eqref{delta.ij},
\begin{align}\label{r}
r(\SZ)	= \frac{\MED^{2}-4\SZ^{2}\log\max_{a,b\in\mcI}\left\{p_{a}/p_{b}\right\}}{2(1+\sqrt{3})\MED},
\end{align}
and 
\begin{align}\label{sigma.zero}
\tau_{ij} = {\MED}\left(2(1+\sqrt{3})\sqrt{N}\left|\log{\left(\frac{\Pxi}{\Pxj}\right)}\right|+4\log\max_{a,b\in\mcI}\left\{\frac{p_{a}}{p_{b}}\right\}\right)^{-\frac{1}{2}}.
\end{align}
\end{lemma}
\begin{IEEEproof}
See Appendix~\ref{Appendix.Lemma.LB}.
\end{IEEEproof}

The results in Lemmas~\ref{Lemma.UB} and \ref{Lemma.LB} can be combined with \eqref{P} to obtain upper and lower bounds on the error probability:
\begin{align}
P(\SZ) & \leq \sumi \Pxi \sumjni h_{ij} \QF\biggl(\frac{\Delta_{ij}(\SZ)}{\SZ}\biggr) \label{P.UBound}
\end{align}
and
\begin{align}
\nonumber
\hspace{-1ex}
P(\SZ) &\ge \sumi \Pxi \sumjnid h_{ij} \left(\QF\Bigl(\frac{\Delta_{ij}(\SZ)}{\SZ}\Bigr)-\QF\Bigl(\frac{\MED}{2\SZ}+\frac{r(\SZ)}{\sqrt{N}\SZ}\Bigr)\right)\cdot\\
&\quad\quad\left(1-2\QF\left(\frac{r(\SZ)}{\sqrt{N}\SZ}\right)\right)^{N-1}, \quad \SZ<\min_{i,j\in\mcI }\tau_{ij}, \label{P.LBound}
\end{align}
where $\Delta_{ij}(\SZ)$, $r(\SZ)$, and $\tau_{ij}$ are given by \eqref{delta.ij}, \eqref{r}, and \eqref{sigma.zero}, respectively. In the next section, it will be proved that both these bounds are tight for asymptotically high SNR.

\section{High-SNR Asymptotics of the SEP and BEP}\label{Sec:Asymptotics}

\subsection{Main Results}\label{Sec:MainResults}
The following theorem gives an asymptotic expression for the error probability in \eqref{P}, i.e., it describes the asymptotic behavior of the MAP and ML detectors, for both SEP and BEP. The results are given in terms of the input probabilities $\Pxi$, the Euclidean distances between constellation points $\deltaij$, the MED of the constellation $\MED$, and the Hamming distances between the binary labels of the constellation points $\hd{i}{j}$. {The results in this section will be discussed later in \secref{Sec:Discussion}. Numerical results will be presented in \secref{Sec:Examples}.}

\begin{theorem}\label{EP.Asym.Theo}
For any input distribution,
\begin{align}\label{EP.Asym}
\limsnszero{
\frac{P(\SZ)}{\QF\bigl(\frac{\MED}{2\SZ}\bigr)}
} = B,
\end{align}
where
\begin{align}\label{EP.B}
B = \sumi \Pxi \sumjnid h_{ij}w_{ij}
\end{align}
and where $h_{ij}$ and $w_{ij}$ are constants given in Table~\ref{Table:EP} and $\QF(\cdot)$ is the Gaussian Q-function.
\end{theorem}
\begin{IEEEproof}
See Appendix~\ref{Appendix.EP.Asym.Theo}.
\end{IEEEproof}

{The following corollary shows that, at high SNR, the ratio between the error probability with MAP and ML detection approaches a constant $R\leq 1$. This constant shows the asymptotic suboptimality of ML detection: when $R<1$ an asymptotic penalty is expected, but when $R=1$ both detectors are asymptotically equivalent.}
\begin{corollary}\label{EP.MAP_vs_ML.Theo}
For any input distribution and for either SEP or BEP,
\begin{align}\label{EP.MAP_vs_ML}
\limsnszero{
\frac{P^{\tnr{map}}(\SZ)}{P^{\tnr{ml}}(\SZ)}
}
=
R,
\end{align}
where
\begin{align}\label{R}
R
=
\frac{B^{\tnr{map}}}{B^{\tnr{ml}}}=
\dfrac{\sumi \Pxi \sumjnid h_{ij}\sqrt{\dfrac{\Pxj}{\Pxi}}}{\sumi \Pxi \sumjnid h_{ij}},
\end{align}
where $h_{ij}$ is a constant given in Table~\ref{Table:EP}. Furthermore, $R\leq 1$ with equality if and only if $\Pxi=\Pxj,\, \forall i,j:\deltaij=\MED$.
\end{corollary}
\begin{IEEEproof}
See Appendix~\ref{Appendix.EP.MAP_vs_ML.Theo}.
\end{IEEEproof}

\subsection{Discussion}\label{Sec:Discussion}

\theoref{EP.Asym.Theo} generalizes \cite[Ths.~3 and 7]{Alvarado12b} to arbitrary multidimensional constellations.\footnote{All the results in \cite{Alvarado12b} are valid for one-dimensional constellations only.} Somewhat surprisingly, \theoref{EP.Asym.Theo} in fact shows that \cite[Ths.~3 and 7]{Alvarado12b} apply verbatim to multidimensional constellations. The result in \theoref{EP.Asym.Theo} for the particular case of SEP with ML detection also coincides with the approximation presented in \cite[Eqs.~(1)--(2)]{Kschischang93}. \theoref{EP.Asym.Theo} can therefore be seen as a formal proof of the asymptotic approximation in \cite[Eqs.~(1)--(2)]{Kschischang93} as well as its generalization to MAP detection for SEP and to both MAP and ML detection for BEP.

Recognizing $B \QF(\MED/(2\SZ))$ as the dominant term in the union bound, \theoref{EP.Asym.Theo} in fact proves that the union bound is tight for both SEP and BEP with arbitrary multidimensional constellations, arbitrary labelings and input distributions, and both MAP and ML detection, which, to the best of our knowledge, has not been previously reported in the literature. The special case of SEP with uniform input distribution and ML detection was elegantly proved in \cite[Eqs.~(7.10)--(7.15)]{Zetterberg77} using an asymptotically tight lower bound.\footnote{An earlier attempt to lower-bound the SEP in the same scenario was presented in \cite[Th.~3]{Gilbert52}, but that bound was incorrect, which can be shown by considering a constellation consisting of three points on a line.}

Note that the lower bound in \eqref{P.LBound} is identical for MAP and ML detection when $R=1$, since $\Delta_{ij}(\SZ)=\MED / 2$ in both cases. The upper bound in \eqref{P.UBound}, however, is always different for MAP and ML detection as long as the symbols are not equally likely, even when $R=1$.

\subsection{Examples}\label{Sec:Examples}

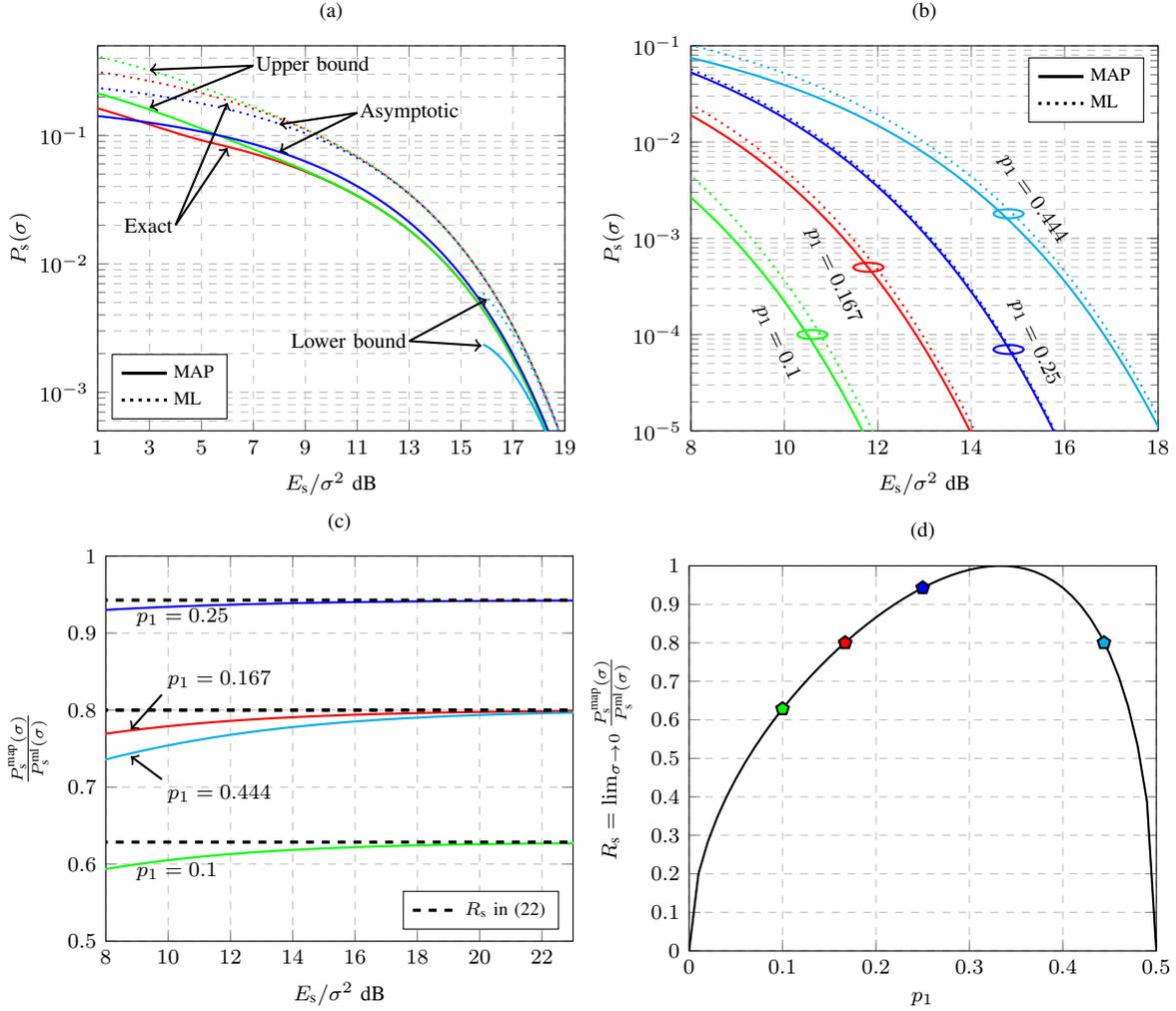
\begin{figure*}
\centering
\begin{tikzpicture}
\pgfplotscreateplotcyclelist{color list}{black,black,red,green,blue,cyan,red,green,blue,cyan}
\begin{semilogyaxis}[cycle list name=color list,
	legend columns=1,
	xminorgrids=true,
    	width=0.43\textwidth,
    	height=0.37\textwidth,
    	grid=both,
    	xmin=1,xmax=19,
    	ymin=5e-4,ymax=0.5,
    	xlabel={$\SX/\SZ^{2}$~dB},
    	xlabel style={yshift=0.1cm},
    	ylabel={$P_{\tnr{s}}(\SZ)$},
    	ylabel style={yshift=-0.2cm},
    	xtick={1,3,...,23},
    	every axis/.append style={font=\footnotesize},
	legend style={legend pos=south west,font=\scriptsize,legend cell align=left},
	grid style={dashed},
	title={(a)}
	]
\addplot[very thick] coordinates {(1e-5,1e-5) (1e-6,1e-6)};\addlegendentry{MAP};
\addplot[very thick,dotted] coordinates {(1e-5,1e-5) (1e-6,1e-6)};\addlegendentry{ML};
\addplot+[thick] file {data/M3_SER_vs_SNR_MAP_exact.dat};
\addplot+[thick] file {data/M3_SER_vs_SNR_MAP_UB.dat};
\addplot+[thick] file {data/M3_SER_vs_SNR_MAP_Asymp.dat};
\addplot+[thick] file {data/M3_SER_vs_SNR_MAP_LB.dat};
\addplot+[thick,dotted] file {data/M3_SER_vs_SNR_ML_exact.dat};
\addplot+[thick,dotted] file {data/M3_SER_vs_SNR_ML_UB.dat};
\addplot+[thick,dotted] file {data/M3_SER_vs_SNR_ML_Asymp.dat};
\addplot+[thick,dotted] file {data/M3_SER_vs_SNR_ML_LB.dat};
\node[coordinate] (A) at (axis cs:3,3.25e-1) {};
\node[coordinate] (B) at (axis cs:3,1.65e-1) {};
\node[anchor=west,align=left,inner sep=0.2ex] (UB) at (axis cs:7,3.5e-1) {Upper bound};
\draw[->,thick] (UB.west) -- (A);
\draw[->,thick] (UB.west) -- (B);
\node[coordinate] (C) at (axis cs:8,7.5e-2) {};
\node[coordinate] (D) at (axis cs:8,1.22e-1) {};
\node[anchor=west,align=left,inner sep=0.2ex] (AS) at (axis cs:11,1.5e-1) {Asymptotic};
\draw[->,thick] (AS.west) -- (C);
\draw[->,thick] (AS.west) -- (D);
\node[coordinate] (E) at (axis cs:6,1.8e-1) {};
\node[coordinate] (F) at (axis cs:6,8e-2) {};
\node[anchor=east,align=left,inner sep=0.2ex] (EX) at (axis cs:4,2e-2) {Exact};
\draw[->,thick] (EX.east) -- (E);
\draw[->,thick] (EX.east) -- (F);
\node[coordinate] (G) at (axis cs:16,5.3e-3) {};
\node[coordinate] (H) at (axis cs:15.8,2.3e-3) {};
\node[anchor=east,align=left,inner sep=0.2ex] (LB) at (axis cs:13,2.5e-3) {Lower bound};
\draw[->,thick] (LB.east) -- (G);
\draw[->,thick] (LB.east) -- (H);
\end{semilogyaxis}
\end{tikzpicture}
\begin{tikzpicture}
\pgfplotscreateplotcyclelist{color list}{black,black,green,red,blue,cyan,green,red,blue,cyan}
\begin{semilogyaxis}[cycle list name=color list,
	legend columns=1,
	xminorgrids=true,
    	width=0.43\textwidth,
    	height=0.37\textwidth,
    	grid=both,
    	xmin=8,xmax=18,
    	ymin=1e-5,ymax=0.1,
    	xlabel={$\SX/\SZ^{2}$~dB},
    	xlabel style={yshift=0.1cm},
    	ylabel={$P_{\tnr{s}}(\SZ)$},
    	ylabel style={yshift=-0.2cm},
    	xtick={8,10,...,18},
    	every axis/.append style={font=\footnotesize},
	legend style={legend pos=north east,font=\scriptsize,legend cell align=left},
	grid style={dashed},
	title={(b)}
	]
\addplot[very thick] coordinates {(1e-5,1e-5) (1e-6,1e-6)};\addlegendentry{MAP};
\addplot[very thick,dotted] coordinates {(1e-5,1e-5) (1e-6,1e-6)};\addlegendentry{ML};
\foreach \i in {0.100,0.167,0.250,0.444}
\addplot+[thick] file {data/Steiner_SER_vs_SNR_MAP_p_\i.dat};
\foreach \i in {0.100,0.167,0.250,0.444}
\addplot+[thick,dotted] file {data/Steiner_SER_vs_SNR_ML_p_\i.dat};
\node[coordinate] (A) at (axis cs:10.6,1e-4) {};
\node[coordinate] (B) at (axis cs:11.8,5e-4) {};
\node[coordinate] (C) at (axis cs:14.8,7e-5) {};
\node[coordinate] (D) at (axis cs:14.8,1.8e-3) {};
\node[rotate=-65,anchor=north] at (axis cs:10.2,1e-4) {$p_{1}=0.1$};
\node[rotate=-65,anchor=north] at (axis cs:11.4,5e-4) {$p_{1}=0.167$};
\node[rotate=-65,anchor=south] at (axis cs:15.0,7e-5) {$p_{1}=0.25$};
\node[rotate=-55,anchor=south] at (axis cs:15.0,2e-3) {$p_{1}=0.444$};
\end{semilogyaxis}
\draw[thick,green,rotate=90] (A) ellipse (0.06 and .2);
\draw[thick,red,rotate=90] (B) ellipse (0.06 and .2);
\draw[thick,blue,rotate=90] (C) ellipse (0.06 and .2);
\draw[thick,cyan,rotate=90] (D) ellipse (0.06 and .2);
\end{tikzpicture}\\
\begin{tikzpicture}
\pgfplotscreateplotcyclelist{color list}{black,green,red,blue,cyan,green,red,blue,cyan}
\begin{axis}[cycle list name=color list,
legend columns=2,
xminorgrids=true,
width=0.43\textwidth,
height=0.37\textwidth,
grid=both,
xmin=8,xmax=23,
ymin=0.5,ymax=1,
xlabel={$\SX/\SZ^{2}$~dB},
xlabel style={yshift=0.1cm},
ylabel={${\frac{P^{\tnr{map}}_{\tnr{s}}(\SZ)}{P^{\tnr{ml}}_{\tnr{s}}(\SZ)}}$},
ylabel style={yshift=-0.2cm},
xtick={0,2,...,23},
ytick={0.5,0.6,0.7,...,1},
every axis/.append style={font=\footnotesize},
legend style={legend pos=south east,font=\scriptsize,legend cell align=left},
grid style={dashed},
title={(c)}
]
\addplot[very thick,dashed,black] coordinates {(1e-5,1e-5) (1e-6,1e-6)};\addlegendentry{$R_{\tnr{s}}$ in \eqref{R}};
\foreach \i in {0.100,0.167,0.250,0.444}
\addplot+[thick] file {data/Steiner_Ratio_vs_SNR_p_\i.dat};
\foreach \i in {0.100,0.167,0.250,0.444}
\addplot+[very thick,dashed,black] file {data/Steiner_Asymptotic_Ratio_vs_SNR_p_\i.dat};
\node[rotate=0,anchor=north west,inner sep=2pt] at (axis cs:8.8,0.94) {$p_{1}=0.25$};
\node[rotate=0,anchor=north west,inner sep=2pt] at (axis cs:8.8,0.61) {$p_{1}=0.1$};
\node[rotate=0,anchor=north west,inner sep=2pt] at (axis cs:9.8,0.71) {$p_{1}=0.444$};
\node[rotate=0,anchor=south west,inner sep=2pt] at (axis cs:9.8,0.82) {$p_{1}=0.167$};
\draw[->,thick] (axis cs:9.8,0.82) -- (axis cs:8.8,0.775);
\draw[->,thick] (axis cs:9.8,0.71) -- (axis cs:8.8,0.74);
\end{axis}
\end{tikzpicture}
\begin{tikzpicture}
\pgfplotscreateplotcyclelist{color list}{red,red,red,red,red,red,red,red}
\begin{axis}[cycle list name=color list,
	legend columns=2,
	xminorgrids=true,
        width=0.43\textwidth,
        height=0.37\textwidth,
        grid=both,
        xmin=0,xmax=0.5,
        ymin=0,ymax=1,
        xlabel={$p_{1}$},
        xlabel style={yshift=0.1cm},
        ylabel={$R_{\tnr{s}}=\limsnszero{\frac{P^{\tnr{map}}_{\tnr{s}}(\SZ)}{P^{\tnr{ml}}_{\tnr{s}}(\SZ)}}$},
        ylabel style={yshift=-0.2cm},
        xtick={0,0.1,0.2,...,0.5},
        ytick={0,0.1,0.2,...,1},
        every axis/.append style={font=\footnotesize},
	legend style={legend pos=south west,font=\scriptsize,legend cell align=left},
	grid style={dashed},
	title={(d)}
	]
\addplot[color=black,thick] table {data/Steiner_ratio_vs_p.dat};
\addplot[color=black,mark=pentagon*,fill=red,thick,only marks,mark size=2.5pt] coordinates {(0.167,0.8)};
\addplot[color=black,mark=pentagon*,fill=cyan,thick,only marks,mark size=2.5pt] coordinates {(0.444,0.8)};
\addplot[color=black,mark=pentagon*,fill=green,thick,only marks,mark size=2.5pt] coordinates {(0.1,0.6285)};
\addplot[color=black,mark=pentagon*,fill=blue,thick,only marks,mark size=2.5pt] coordinates {(0.25,0.9428)};
\end{axis}
\end{tikzpicture}
\caption{Results obtained for the constellations in Example~\ref{example.1D}: {(a) SEP and bounds with MAP and ML detection for the asymmetric constellation}, (b) SEP with MAP and ML detection {for the symmetric constellation}, (c) ratio of SEPs and asymptote {for the symmetric constellation} given by \cororef{EP.MAP_vs_ML.Theo}, and (d) asymptote {for the symmetric constellation} as a function of the symbol probability $p_{1}$.}
\label{Steiner94}
\end{figure*}

\begin{example}\label{example.1D}
{
Consider the one-dimensional \emph{asymmetric} constellation with $M=3$, $(x_1,x_2,x_3) = (-1,0,+2)$, and $(p_1,p_2,p_3) = (0.62,0.07,0.31)$. This probability distribution is chosen so that the average of the constellation is zero. \figref{Steiner94}~(a) shows the exact SEP with MAP and ML detection, which can be analytically calculated, together with the upper bounds in \eqref{P.UBound} (green), the lower bounds in  \eqref{P.LBound} (cyan), and the asymptotic approximations $P_{\tnr{s}}(\SZ)\approx B_{\tnr{s}}\QF(\MED/(2\SZ))$ from \eqref{EP.Asym} (blue) with 
$B^{\tnr{map}}_{\tnr{s}}=2\sqrt{p_1 p_2}=0.4167$ and 
$B^{\tnr{ml}}_{\tnr{s}}=p_1+p_2=0.6900$, i.e., 
$R_{\tnr{s}}=0.6038$, given by \cororef{EP.MAP_vs_ML.Theo}. The solid and dotted curves represent MAP and ML detection, respectively. The lower bounds are only defined when $\SX/\SZ^{2}>15.8$ dB, due to the restrictions on $\SZ$ in \eqref{P.LBound}. In this example the asymptotic approximation for the ML detector is below the exact SEP, while for the MAP detector the asymptotic approximation is above the exact SEP when $\SX/\SZ^{2}>2.6$ dB. \figref{Steiner94}~(a) also shows that there is a difference in the SEP between the MAP and ML detector and that the upper and lower bounds are tight and converge to both the exact SEP and the asymptotic approximation for high SNR. 
}

{Now consider instead the one-dimensional \emph{symmetric}} constellation with $M=3$, $(x_1,x_2,x_3) = (-1,0,+1)$, and $(p_1,p_2,p_3) = (p_1,1-2p_1,p_1)$, where $0<p_{1}<1/2$. If $p_{1}=1/3$, an equally likely and equally spaced $3$-ary constellation is obtained. If $p_1 = 1/K$, this constellation is equivalent to a constellation with $K$ equally likely points, of which $K-2$ are located at the origin; such a constellation was used in \cite{Steiner94} to disprove the so-called strong simplex conjecture.

In \figref{Steiner94}~(b), the exact SEP with MAP and ML detection, which can be analytically calculated, is shown for different values of $p_{1}$.
There is a clear performance differences between the two detectors when $p_{1}\neq 1/3$. According to \cororef{EP.MAP_vs_ML.Theo}, $B^{\tnr{map}}_{\tnr{s}}=4\sqrt{p_1(1-2p_1)}$ and $B^{\tnr{ml}}_{\tnr{s}}=2(1-p_1)$, i.e., $R_{\tnr{s}}=2\sqrt{p_1(1-2p_1)}/(1-p_1)$. \figref{Steiner94}~(c) shows the ratio of the SEP curves and how these converge to $R_{\tnr{s}}$ as $\SZ\rightarrow0$ (indicated by the horizontal dashed lines). 
For $p_{1}=0.167$ and $p_{1}=0.444$, the asymptote is the same ($R_{\tnr{s}}=0.8$); however, their SEP performance is quite different (see \figref{Steiner94}~(b)). This can be explained using the results in \figref{Steiner94}~(d), where the solid line shows $R_{\tnr{s}}$. 
The two markers when $R_{\tnr{s}}=0.8$ correspond to $p_{1}=0.167$ and $p_{1}=0.444$, which explains the results in \figref{Steiner94}~(c) for those values of $p_{1}$.
\end{example}

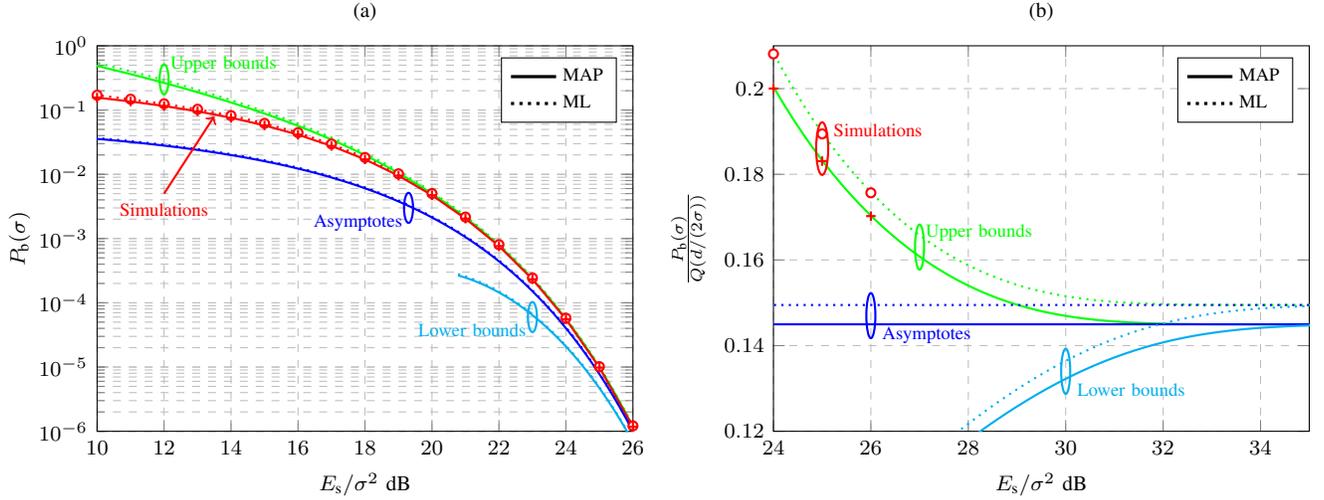
\begin{figure*}
\centering
\begin{tikzpicture}
\pgfplotscreateplotcyclelist{color list}{black,black,green,red,blue,cyan,green,red,blue,cyan}
\begin{semilogyaxis}[cycle list name=color list,
	legend columns=1,
	xminorgrids=true,
    	width=0.48\textwidth,
    	height=0.37\textwidth,
    	grid=both,
    	xmin=10,xmax=26,
    	ymin=1e-6,ymax=1,
    	xlabel={$\SX/\SZ^{2}$~dB},
    	xlabel style={yshift=0.1cm},
    	ylabel={$P_{\tnr{b}}(\SZ)$},
    	ylabel style={yshift=-0.2cm},
    	every axis/.append style={font=\footnotesize},
	legend style={legend pos=north east,font=\scriptsize,legend cell align=left},
	grid style={dashed},
	title={(a)}
	]
\addplot[very thick] coordinates {(1e-5,1e-5) (1e-6,1e-6)};\addlegendentry{MAP};
\addplot[very thick,dotted] coordinates {(1e-5,1e-5) (1e-6,1e-6)};\addlegendentry{ML};
\addplot+[thick,dotted] file {data/Valenti_BEP_ML_UB.txt};
\addplot+[thick,dotted,mark=*,mark size=1.7,mark options={fill=white,solid}] file {data/Valenti_BEP_ML_Sim.txt};
\addplot+[thick,dotted] file {data/Valenti_BEP_ML_Asym.txt};
\addplot+[thick,dotted] file {data/Valenti_newBEP_ML_LB.txt}; 
\addplot+[thick] file {data/Valenti_BEP_MAP_UB.txt};
\addplot+[thick,mark=+,mark size=1.8,mark options={fill=white}] file {data/Valenti_BEP_MAP_Sim.txt};
\addplot+[thick] file {data/Valenti_BEP_MAP_Asym.txt};
\addplot+[thick] file {data/Valenti_newBEP_MAP_LB.txt}; 
\node[coordinate] (EL1) at (axis cs:12,0.3) {};
\node[coordinate] (EL3) at (axis cs:19.3,3e-3) {};
\node[coordinate] (EL4) at (axis cs:23,6e-5) {};
\draw[->,thick,red] (axis cs:12,5e-3)node[red,anchor=north]{{\scriptsize{Simulations}}} -- (axis cs:13.5,8e-2);
\end{semilogyaxis}
\draw[thick,green,rotate=0] (EL1) ellipse (0.06 and .2) node[anchor=south west,inner sep=2pt]{{\scriptsize{Upper bounds}}};
\draw[thick,blue,rotate=0] (EL3) ellipse (0.06 and .2) node[anchor=north east,inner sep=2pt]{{\scriptsize{Asymptotes}}};
\draw[thick,cyan,rotate=0] (EL4) ellipse (0.06 and .2) node[anchor=north east,inner sep=2pt]{{\scriptsize{Lower bounds}}};
\end{tikzpicture}
\begin{tikzpicture}
\pgfplotscreateplotcyclelist{color list}{black,black,green,red,blue,cyan,green,red,blue,cyan}
\begin{axis}[cycle list name=color list,
	legend columns=1,
	xminorgrids=true,
	width=0.48\textwidth,
    	height=0.37\textwidth,
	grid=both,
	xmin=24,xmax=35,
	ymin=0.12,ymax=0.21,
	xlabel={$\SX/\SZ^{2}$~dB},
	xlabel style={yshift=0.1cm},
	ylabel={$\frac{P_{\tnr{b}}(\SZ)}{\QF(\MED/(2\SZ))}$},
	ylabel style={yshift=-0.1cm},
	every axis/.append style={font=\footnotesize},
	legend style={legend pos=north east,font=\scriptsize,legend cell align=left},
	grid style={dashed},
	title={(b)}
]
\addplot[very thick] coordinates {(1e-5,1e-5) (1e-6,1e-6)};\addlegendentry{MAP};
\addplot[very thick,dotted] coordinates {(1e-5,1e-5) (1e-6,1e-6)};\addlegendentry{ML};
\addplot+[thick] file {data/Valenti_BEPQ_MAP_UB.txt};
\addplot+[thick,only marks,mark=+,mark size=1.8,mark options={fill=white}] file {data/Valenti_BEPQ_MAP_Sim.txt};
\addplot+[thick] file {data/Valenti_BEPQ_MAP_Asym.txt};
\addplot+[thick] file {data/Valenti_BEPQ_MAP_LB.txt};
\addplot+[thick,dotted] file {data/Valenti_BEPQ_ML_UB.txt};
\addplot+[thick,only marks,dotted,mark=*,mark size=1.7,mark options={fill=white,solid}] file {data/Valenti_BEPQ_ML_Sim.txt};
\addplot+[thick,dotted] file {data/Valenti_BEPQ_ML_Asym.txt};
\addplot+[thick,dotted] file {data/Valenti_BEPQ_ML_LB.txt};
\node[coordinate] (ER1) at (axis cs:27,0.163) {};
\node[coordinate] (ER2) at (axis cs:25,0.186) {};
\node[coordinate] (ER3) at (axis cs:26,0.147) {};
\node[coordinate] (ER4) at (axis cs:30,0.134) {};
\end{axis}
\draw[thick,green,rotate=0] (ER1) ellipse (0.06 and .3) node[anchor=south west,inner sep=2pt]{{\scriptsize{Upper bounds}}};
\draw[thick,red,rotate=0] (ER2) ellipse (0.08 and .35) node[anchor=south west,inner sep=4pt]{{\scriptsize{Simulations}}};
\draw[thick,blue,rotate=0] (ER3) ellipse (0.06 and .3) node[anchor=north west,inner sep=4pt]{{\scriptsize{Asymptotes}}};
\draw[thick,cyan,rotate=0] (ER4) ellipse (0.06 and .3) node[anchor=north west,inner sep=4pt]{{\scriptsize{Lower bounds}}};
\end{tikzpicture}
\caption{Results obtained for the constellation in Example~\ref{Example.Valenti}: (a) BEP with MAP and ML detection, and (b) ratio between BEPs in (a) and $\QF(\MED/(2\SZ))$.}
\label{Valenti12}
\end{figure*}

\begin{example}\label{Example.Valenti}
Consider again the constellation in Example~\ref{Example.Regions} (see \figref{Valenti_32APSK}) with the labeling specified in \cite[Fig.~2]{Valenti12}. \figref{Valenti12}~(a) shows the simulated BEPs (red markers) together with the upper bounds in \eqref{P.UBound} (green), the lower bounds in  \eqref{P.LBound} (cyan), and the asymptotic approximations $P_{\tnr{b}}(\SZ)\approx B_{\tnr{b}}\QF(\MED/(2\SZ))$ from \eqref{EP.Asym} (blue). The solid and dotted curves represent MAP and ML detection, respectively. The lower bounds are only defined when $\SX/\SZ^{2}>20.78$ dB, due to the restrictions on $\SZ$ in \eqref{P.LBound}. These result show very small differences between the MAP and ML detectors. To see the asymptotic behavior more clearly, \figref{Valenti12}~(b) shows the ratio between the eight curves in \figref{Valenti12}~(a) and $\QF(\MED/(2\SZ))$. It is clear that the simulated BEPs closely follow the upper bounds at these SNR values. These results also show that both the upper and lower bounds converge to $B^{\tnr{map}}_{\tnr{b}}=0.1450$ and $B^{\tnr{ml}}_{\tnr{b}}=0.1495$ for MAP and ML detection, respectively, as predicted by \theoref{EP.Asym.Theo}. Unlike \figref{Valenti12}~(a), \figref{Valenti12}~(b) clearly shows the asymptotic difference between the MAP and ML detectors, {since $R_{\tnr{b}}=0.97$}.

{
The gap between $P_{\tnr{b}}^{\tnr{map}}$ and $P_{\tnr{b}}^{\tnr{ml}}$ depends on the bit labeling, but not as strongly as on the probability distribution. It can be shown that for the probabilities in this example, $0.956 < R_{\tnr{b}} < 0.989$ for any labeling. On the other hand, for this labeling, $R_{\tnr{b}}$ can be made equal to any value in the interval $(0,1]$ by changing the probabilities.
}
\end{example}

\cororef{EP.MAP_vs_ML.Theo} gives necessary and sufficient conditions for the asymptotic optimality of ML detection for both SEP and BEP. A nonuniform distribution will in general give $R<1$, although there are exceptions. Consider for example a constellation that can be divided into \emph{clusters}, where all pairs of constellation points in different clusters are at distances larger than the MED. Then ML detection is asymptotically optimal (i.e., $R=1$) if the probabilities of all constellations points \emph{within} a cluster are equal, even if the clusters have different probabilities. In this special case, \eqref{EP.B} for SEP yields $B^{\tnr{map}}_{\tnr{s}}=B^{\tnr{ml}}_{\tnr{s}}=\sumi p_i G_i$, where $G_i$ is the number of neighbors at MED from point $i$. We illustrate this concept with the following example.

\begin{figure}
\centering
\raisebox{1ex}{
\hspace{6ex}
\begin{tikzpicture}
\pgfplotscreateplotcyclelist{color list}{black,black,green,red,blue,cyan,green,red,blue,cyan}
\begin{axis}[
	axis lines=none,
    	width=0.22\textwidth,
    	height=0.22\textwidth,
    	grid=both,
    	xmin=-4,xmax=4,
    	ymin=-4,ymax=4,
    	every axis/.append style={font=\footnotesize},
	legend style={legend pos=north east,font=\scriptsize,legend cell align=left},
	grid style={dashed},
	title={(a)}
	]
	\addplot+[black,thick,mark=*,mark options={fill=black},mark size=4.69pt] file {data/dodeca2_16_inner_ring.txt};
	\addplot+[black,thick,mark=*,mark options={fill=black},mark size=1pt] file {data/dodeca2_16_outer_ring.txt};
\end{axis}
\end{tikzpicture}
}
\begin{tikzpicture}
\pgfplotscreateplotcyclelist{color list}{black,black,green,red,blue,cyan,green,red,blue,cyan}
\begin{axis}[cycle list name=color list,
	legend columns=1,
	xminorgrids=true,
    	width=0.48\textwidth,
        	height=0.37\textwidth,
    	grid=both,
    	xmin=0,xmax=30,
    	ymin=1.8,ymax=2.8,
    	xlabel={$\SX/\SZ^{2}$~dB},
    	xlabel style={yshift=0.1cm},
	ylabel={$\frac{P_{\tnr{s}}(\SZ)}{\QF(\MED/(2\SZ))}$},
	ylabel style={yshift=-0.2cm},
    	every axis/.append style={font=\footnotesize},
	legend style={legend pos=north east,font=\scriptsize,legend cell align=left},
	grid style={dashed},
	title={(b)}
	]
\addplot[very thick] coordinates {(1e-5,1e-5) (1e-6,1e-6)};\addlegendentry{MAP};
\addplot[very thick,dotted] coordinates {(1e-5,1e-5) (1e-6,1e-6)};\addlegendentry{ML};
\addplot+[thick] file {data/APSK-4-12_SEPQ_MAP_UB.txt};
\addplot+[thick,only marks,mark=+,mark size=1.8,mark options={fill=white}] file {data/APSK-4-12_SEPQ_MAP_Sim.txt};
\addplot+[thick] file {data/APSK-4-12_SEPQ_MAP_Asym.txt};
\addplot+[thick] file {data/APSK-4-12_SEPQ_MAP_LB.txt};
\addplot+[thick,dotted] file {data/APSK-4-12_SEPQ_ML_UB.txt};
\addplot+[thick,only marks,dotted,mark=*,mark size=1.7,mark options={fill=white,solid}] file {data/APSK-4-12_SEPQ_ML_Sim.txt};
\addplot+[thick,dotted] file {data/APSK-4-12_SEPQ_ML_Asym.txt};
\addplot+[thick,dotted] file {data/APSK-4-12_SEPQ_ML_LB.txt};
\node[coordinate] (T1) at (axis cs:9.8,2.77) {};
\node[blue,anchor=north] (T2) at (axis cs:9,2) {{\scriptsize{Asymptotes}}};
\node[cyan,anchor=east] (T3) at (axis cs:23.5,1.88) {{\scriptsize{Lower bounds}}};
\draw[->,thick,red] (axis cs:5,2.4) node[red,anchor=north]{{\scriptsize{Simulations}}} -- (axis cs:7.9,2.72);
\draw[->,thick,red] (axis cs:5,2.31) -- (axis cs:7.9,2.199);
\end{axis}
\draw[thick,green,rotate=90] (T1) ellipse (0.06 and .5) node[anchor=west,inner sep=17pt]{{\scriptsize{Upper bounds}}};
\end{tikzpicture}
\caption{Results obtained for the constellation in Example~\ref{example.2D}: (a) Constellation where the pairs of symbols at MED are marked with solid lines and the symbol probabilities are indicated by the point areas, and (b) asymptotic performance shown as the ratio between SEPs and $\QF(\MED/(2\SZ))$.}
\label{Thomas74}
\end{figure}
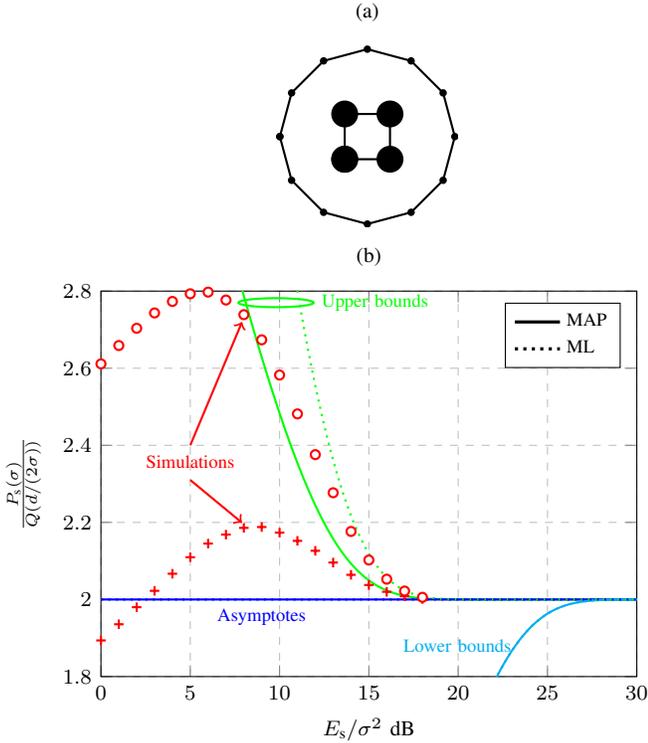

\begin{example}\label{example.2D}
Fig.~\ref{Thomas74}~(a) illustrates the two-dimensional constellation in \cite[Fig.~3 (d)]{Thomas74}. We let the symbols in the inner ring be used with probability $p_{1}$ each and the symbols in the outer ring with probability $p_{2}=(1-4p_{1})/12$. {The radii of the two rings are $r_1 = 0.71d$ and $r_2 = 1.93 d$, and the average symbol energy is $\SX = 4 p_1 r_1^2 + 12 p_2 r_2^2$.} Fig.~\ref{Thomas74}~(b) shows the simulated ratio $P_{\tnr{s}}(\SZ)/\QF(\MED/(2\SZ))$ when $p_1=0.22$ and $p_2=0.01$ for ML (red circles) and MAP (red crosses) detection.
The upper bounds in \eqref{P.UBound}, the lower bounds in  \eqref{P.LBound}, and the asymptotic expression, all divided by $\QF(\MED/(2\SZ))$, are included as green, cyan, and blue curves, respectively. In this case, the lower bounds for ML and MAP detection are identical, as are the asymptotes. For this specific constellation, $G_i=2$ for all $i\in\mcIX$, and hence, $B^{\tnr{map}}_{\tnr{s}}=B^{\tnr{ml}}_{\tnr{s}}=2$, independently of $p_1$ and $p_2$, which implies $R_{\tnr{s}}=1$. {In terms of BEP, $B^{\tnr{map}}_{\tnr{b}}=B^{\tnr{ml}}_{\tnr{b}}$, hence $R_{\tnr{b}}=1$, regardless of both the labeling and $p_1$ and $p_2$. This shows that for this particular nonuniform constellation, both ML and MAP detectors are asymptotically equivalent for SEP and BEP for all labelings.}
\end{example}

\section{Conclusions}\label{Sec:Conclusions}

In this paper, an analytical characterization of the asymptotic behavior of the MAP and ML detectors in terms of SEP and BEP for arbitrary multidimensional constellations over the AWGN channel was presented. The four obtained results from \theoref{EP.Asym.Theo} and Table~\ref{Table:EP} can be summarized as
\begin{align}
\label{BEP.map.app}
P_{\tnr{b}}^{\tnr{map}}(\SZ) &\approx \QF\biggl(\frac{\MED}{2\SZ}\biggr) \sumijnid \dfrac{\hd{i}{j}}{m}\sqrt{\Pxi\Pxj},\\
\label{BEP.ml.app}
P_{\tnr{b}}^{\tnr{ml}}(\SZ) &\approx \QF\biggl(\frac{\MED}{2\SZ}\biggr) \sumijnid \dfrac{\hd{i}{j}}{m} \Pxi,\\
\label{SEP.map.app}
P_{\tnr{s}}^{\tnr{map}}(\SZ) &\approx \QF\biggl(\frac{\MED}{2\SZ}\biggr) \sumijnid \sqrt{\Pxi\Pxj},\\
\label{SEP.ml.app}
P_{\tnr{s}}^{\tnr{{ml}}}(\SZ) &\approx \QF\biggl(\frac{\MED}{2\SZ}\biggr) \sumijnid \Pxi,
\end{align}
where the relative error in all approximations approaches zero as $\SZ\rightarrow0$. The expressions for MAP and ML are equal if and only if $\Pxi=\Pxj,\, \forall i,j:\deltaij=\MED$.

Somewhat surprisingly, the results in this paper are the first ones that address the problem in such a general setup. The theoretical analysis shows that for nonuniform input distributions, ML detection is in general asymptotically suboptimal. In most practically relevant cases, however, MAP and ML detection give very similar asymptotic results. The results in this paper are first-order only. An asymptotic analysis considering higher order terms is left for further investigation. 

{Most modern transceivers based on high-order modulation formats use a receiver that operates at a bit level (i.e., a bit-wise receiver). Because of this, the binary labeling of the constellation plays a fundamental role in the system design. Furthermore, the use of nonequally likely symbols (probabilistic shaping) has recently received renewed attention in the literature. In this context, the asymptotic BER expressions presented in this paper can be used to optimize both the constellations and the binary labeling. This is also left for further investigation.}

\section*{Acknowledgment}\label{Sec:Ack}

We would like to thank Prof.~F.~R.~Kschischang for insightful comments on our poster ``Asymptotic Error Probability Expressions for the MAP Detector and Multidimensional Constellations,'' presented at the Recent Results Poster Session of ISIT 2016, Barcelona.

\appendices

\section{Proof of \lemmaref{Lemma.UB}}\label{Appendix.Lemma.UB}
\begin{figure}[t]
\scalebox{0.9}{
\centerline{
\footnotesize{
\hspace{-3.5ex}
\begin{tikzpicture}[tight background]
\draw [lightgray,fill=lightgray] plot [smooth] coordinates {(-100pt,45pt) (-110pt,130pt) (-20pt,110pt) (30pt,130pt) (70pt,115pt) (130pt,130pt) (100pt,45pt)};
\draw [black,fill=lightblue,opacity=0.5] plot coordinates {(-50pt,45pt) (-70pt,100pt) (80pt,110pt) (50pt,45pt)};
\draw [black,fill=lightred,opacity=0.5] plot coordinates {(50pt,45pt) (70pt,2.5pt) (130pt,-15pt) (130pt,90pt) (80pt,110pt) };
\draw [black,fill=lightgreen,opacity=0.5] plot coordinates {(70pt,2.5pt) (20pt,-56pt) (80pt,-90pt) (120pt,-70pt) (130pt,-15pt)};
\node[] at (55pt,95pt) {$\mcR_{j}^{\tnr{map}}(\SZ)$};
\node[] at (110pt,5pt) {$\mcR_{k}^{\tnr{map}}(\SZ)$};
\node[] at (100pt,-50pt) {$\mcR_{n}^{\tnr{map}}(\SZ)$};
\node[] at (-90pt,110pt) {$\mcH_{ij}^{\tnr{map}}(\SZ)$};
\fill[thick] (0pt,0pt) circle (2pt) node[anchor=south east] {$\bx_{i}$};
\draw[black,thick,dashed] (0pt,0pt) circle (80pt);
\draw[thick,dashed] (0pt,80pt) circle (80pt);
\draw[<->,thick] (0.5pt,81.5pt) -- node[anchor=east] {$\MED$} +(55:78pt) ;
\fill[thick,black] (0pt,80pt) circle (2pt) node[anchor=south east] {$\bx_{j}$};
\fill[thick,black] (90pt,40pt) circle (2pt) node[anchor=north west] {$\bx_{k}$};
\draw[<->,thick] (-120:2pt) -- node[anchor=west] {$\MED$} (-120:80pt) ;
\draw[black,thick] (-10pt,30pt) rectangle (10pt,50pt);
\draw[black,thick,fill=gray] (-10pt,45pt) rectangle (10pt,50pt);
\draw[black,thick,dashed] (0pt,40pt) circle (14.1pt);
\draw[->,thick,black] (0pt,2pt) -- +(90:76pt) node[anchor=north west] {$\bdij$};
\draw[<->,thick] (0pt,40pt) -- node[anchor=south east,inner sep=-0.5pt] {$r$} (-10pt,30pt); 
\draw[-,black,thick] (-100pt,45pt) -- (100pt,45pt) ;
\draw[-,thick,black] (-2pt,40pt) --  (2pt,40pt) {};
\draw[<-,thick] (2.2pt,40pt) --  (30pt,35pt) node[anchor=west,inner sep=1pt] {$\ov{\bx}_{ij}$};
\draw[<-,thick] (-5pt,47.5pt) --  (-65pt,-10pt) node[anchor=north,inner sep=1pt] {Orthotope $\mcR_{j}^{\tnr{map}}(\SZ)\cap\Cij(\SZ)$};
\draw[<->,thick] (-100pt,0pt) -- node[anchor=east] {$|\Delta_{ij}(\SZ)|$} (-100pt,45pt) ;
\draw[-,thick,dashed] (-100pt,0pt) -- (0pt,0pt) ;
\fill[thick,black] (-40:80pt) circle (2pt) node[anchor=north west] {$\bx_{n}$};
\draw[black,thick,dashed] (-40:40pt) circle (14.1pt);
\draw[black,thick,rotate=-40,fill=gray] (30pt,-10pt) rectangle (50pt,10pt);
\draw[<-,thick] (30pt,-20pt) -- (35pt,-5pt) node[anchor=south west,inner sep=0pt] {$\Cin(\SZ)$};
\draw[-,thick,black,rotate=-40] (40pt,-2pt) --  (40pt,2pt) {};
\draw[->,thick] (-40:2pt) -- (-40:35pt) -- node[anchor=west] {$\bdin$} (-40:78pt);
\end{tikzpicture}}}
}
\caption{Geometric representation of the proofs of Lemma~\ref{Lemma.UB} and \ref{Lemma.LB} for a 2D constellation with MAP detection. The MAP decision regions are shown for $\bx_{j}$, $\bx_{k}$, and $\bx_{n}$, the half-space in \eqref{sep.i.map.halfspace} for $\bx_{j}$ ($\mcH_{ij}^{\tnr{map}}(\SZ)$), and the hypercube for $\bx_{n}$ ($\Cin(\SZ)$).}
\label{proof_sketch_2D}
\end{figure}
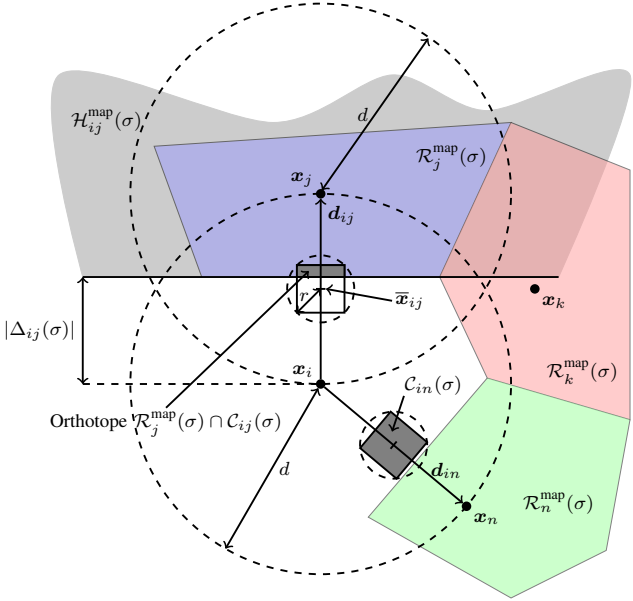

For MAP detection, we have from \eqref{Fij.2} and \eqref{map.j.region} that
\begin{align}\label{Fij.UB.1}
\tpMAP{i}{j}(\SZ) \leq \Pr\bigl\{\bY \in \mcH_{ij}^{\tnr{map}}(\SZ)|\bX=\bx_{i}\bigr\},
\end{align}
where $\mcH_{ij}^{\tnr{map}}(\SZ)$ is the half-space determined by a \emph{pairwise} MAP decision (see \eqref{map}--\eqref{map.j.region}), i.e.,
\begin{align}\label{sep.i.map.halfspace.def}
\mcH_{ij}^{\tnr{map}}(\SZ) &= \set{\by\in\mathbb{R}^{N}: \Pxi\pdf(\by|\bx_{i}) \leq \Pxj\pdf(\by|\bx_{j})}.
\end{align}
Using \eqref{pdf.channel}, \eqref{sep.i.map.halfspace.def} can be expressed as
\begin{align}
\mcH_{ij}^{\tnr{map}}(\SZ)
\label{sep.i.map.halfspace}
& = \biggl\{\by\in\mathbb{R}^{N}:\inner{\by-\bx_{i}}{\frac{\bdij}{\deltaij}} \geq \Delta_{ij}(\SZ)\biggr\},
\end{align}
where $\Delta_{ij}(\SZ)$ is given by \eqref{delta.ij} and $\bdij= \bx_{j}-\bx_{i}$. The value of $|\Delta_{ij}(\SZ)|$ is the shortest Euclidean distance between $\bx_{i}$ and the hyperplane defining the half-space $\mcH_{ij}^{\tnr{map}}(\SZ)$. For a geometric interpretation, see $\mcH_{ij}^{\tnr{map}}(\SZ)$ in \figref{proof_sketch_2D}.

Using \eqref{sep.i.map.halfspace}, \eqref{Fij.UB.1} can be calculated as
\begin{align}\label{Fij.UB.2}
\tpMAP{i}{j}(\SZ) 	& \leq  
\Pr\left\{ \inner{\bY-\bX}{\frac{\bdij}{\deltaij}} \ge \Delta_{ij}(\SZ) \mid \bX = \bx_i \right\} \\
\label{Fij.UB.3}
			& = \QF\biggl(\frac{\Delta_{ij}(\SZ)}{\SZ}\biggr),
\end{align}
where \eqref{Fij.UB.3} follows from \eqref{AWGN}--\eqref{pdf.channel} by recognizing $\inner{\bY-\bX}{\bdij/\deltaij}$ as a zero-mean Gaussian random variable with variance $\SZ^2$.

The proof for the ML case is analogous but starts from \eqref{ml.j.region} instead of \eqref{map.j.region}. If follows straightforwardly that $\mcH_{ij}^{\tnr{ml}}(\SZ)$ is also given by \eqref{sep.i.map.halfspace}, where now $\Delta_{ij}(\SZ)={\deltaij}/{2}$ defines a hyperplane half-way between $\bx_{i}$ and $\bx_{j}$.

\section{Proof of \lemmaref{Lemma.LB}}\label{Appendix.Lemma.LB}

To lower-bound \eqref{Fij.2} for MAP detection, we first ignore all the contributions of constellation points not at MED, i.e., we use $\Pr\Set{\bY \in \mcR_{j}^{\tnr{map}}(\SZ)|\bX=\bx_{i}}\geq 0$, for all $j$ such that $\deltaij>\MED$. This gives the first case in \eqref{Lemma.LB.eq}.

The case of $\deltaij=\MED$ is addressed by first defining an $N$-dimensional hypersphere centered at the mid-point $\ov{\bx}_{ij}= \frac{1}{2}(\bx_{i}+\bx_{j})$ and of radius $r(\SZ)$ defined in \eqref{r}. Second, we inscribe an $N$-dimensional hypercube with half-side $r(\SZ)/\sqrt{N}$ inside this hypersphere. And third, we rotate this hypercube so that one of its sides is perpendicular to $\bdij$. We denote this hypercube by $\Cij(\SZ)$. For a geometric interpretation when $N=2$, see \figref{proof_sketch_2D}. 

To lower-bound \eqref{Fij.2}, we integrate $\pdf(\by|\bx)$ in \eqref{pdf.channel} over the intersection of the MAP region $\mcR_{j}^{\tnr{map}}(\SZ)$ and the hypercube $\Cij(\SZ)$, i.e.,
\begin{align}\label{Lemma.LB.1}
\tpMAP{i}{j}(\SZ) 	& = \int_{\mcR_{j}^{\tnr{map}}(\SZ)} \pdf(\by|\bx_{i})\, \tnr{d}\by\\
\label{Lemma.LB.2}
			& \geq \int_{\mcR_{j}^{\tnr{map}}(\SZ)\,\cap\,\Cij(\SZ)} \pdf(\by|\bx_{i})\, \tnr{d}\by.
\end{align}
We will prove that for sufficiently low values of $\SZ$, the integration region in \eqref{Lemma.LB.2} is an orthotope (hyperrectangle), as illustrated for $\bx_{i}$ and $\bx_{j}$ in \figref{proof_sketch_2D}. This will be done in three steps. We will show first that the hypercube $\Cij(\SZ)$ is nonempty, second that it does not intersect any region $\mcR_{k}^{\tnr{map}}(\SZ)$ for $k \notin \set{i,j}$, and third that $\Cij(\SZ)$ intersects both $\mcR_{i}^{\tnr{map}}(\SZ)$ and $\mcR_{j}^{\tnr{map}}(\SZ)$. Together these three facts imply that $\mcR_{j}^{\tnr{map}}(\SZ)\cap\Cij(\SZ)$ is an orthotope, whose dimensions are then determined, which allows the integral in \eqref{Lemma.LB.2} to be calculated exactly.

For the first step, combining \eqref{r} and \eqref{sigma.zero} and rearranging terms yields for any $i,j\in\mcIX$ with $j\neq i$
\begin{align}\label{d2diff}
\frac{\MED^2}{\SZ^2} - \frac{\MED^2}{\tau_{ij}^2}
= 2(1+\sqrt{3}) \left(\frac{\MED r(\SZ)}{\SZ^2} - \sqrt{N}\left|\log{\frac{\Pxi}{\Pxj}}\right| \right).
\end{align}
If $\SZ < \tau_{ij}$, then \eqref{d2diff} implies
\begin{align}\label{r.ineq}
\frac{r(\SZ)}{\sqrt{N}} &> \frac{\SZ^2}{\MED}\left|\log{\frac{\Pxi}{\Pxj}}\right|.
\end{align}
This inequality, to which we will return later, shows that $r(\SZ) > 0$ and hence that $\Cij(\SZ)$ is not empty.

For the second step, we consider any $i,j,k\in\mcIX$ such that $\deltaij = \MED$ and $k \notin \set{i,j}$. We have
\begin{align}
\MED^2 &\le \min_{\ell \in \set{i,j}} \|\bx_k - \bx_\ell\|^2 \\
& = \min_{\ell \in \set{i,j}}  \|(\bx_k - \ov{\bx}_{ij}) - (\bx_\ell - \ov{\bx}_{ij})\|^2 \\
& = \|\bx_k - \ov{\bx}_{ij}\|^2 - 2\max_{\ell \in \set{i,j}} \inner{\bx_k - \ov{\bx}_{ij}}{\bx_\ell - \ov{\bx}_{ij}}+ \frac{\MED^2}{4} \label{xlbound1} \\
& \le \|\bx_k - \ov{\bx}_{ij}\|^2 + \frac{\MED^2}{4}, \label{xlbound2} 
\end{align}
where \eqref{xlbound1} follows because $\|\bx_i - \ov{\bx}_{ij}\|^2 = \|\bx_j - \ov{\bx}_{ij}\|^2 = \MED^2/4$ and \eqref{xlbound2} because $\bx_i - \ov{\bx}_{ij} = - (\bx_j - \ov{\bx}_{ij})$ in the second term of \eqref{xlbound1}. Hence, $\|\bx_k-\ov{\bx}_{ij}\| \ge \MED\sqrt{3}/2$.
Consider now any point $\by \in \Cij(\SZ)$. By the triangle inequality, $\|\by-\bx_k\| \ge \|\bx_k-\ov{\bx}_{ij}\| - \|\by-\ov{\bx}_{ij}\| \ge \MED\sqrt{3}/2-r(\SZ)$ and $\|\by-\bx_i\| \le \|\bx_i-\ov{\bx}_{ij}\| + \|\by-\ov{\bx}_{ij}\| \le \MED/2+r(\SZ)$,
which are then combined into
\begin{align}
\label{empty.eq.2}
\|\by-\bx_{k}\|^{2}-\|\by-\bx_{i}\|^{2}	& \geq \left(\frac{\MED\sqrt{3}}{2}-r(\SZ)\right)^{2}-\left(\frac{\MED}{2}+r(\SZ)\right)^{2}\\
&= \frac{\MED^2}{2} - \left(1+\sqrt{3}\right) \MED r(\SZ) \\
\label{empty.eq.3}
& = 2\SZ^{2}\log\max_{a,b\in\mcI}\left\{\frac{p_{a}}{p_{b}}\right\} \\
\label{empty.eq.4}
& \ge 2\SZ^{2}\log\frac{p_k}{p_i},
\end{align}
where \eqref{empty.eq.3} follows from \eqref{r}. Rearranging terms,
\begin{align} \label{map.ineq}
p_k \explow{-\frac{\|\by-\bx_{k}\|^2}{2\SZ^2}} \le p_i \explow{-\frac{\|\by-\bx_{i}\|^2}{2\SZ^2}},
\end{align}
which via \eqref{pdf.channel} and \eqref{map.j.region} implies $\by \notin \mcR_{k}^{\tnr{map}}(\SZ).$\footnote{If \eqref{map.ineq} is an equality, $\by$ may lie on the boundary of $\mcR_{k}^{\tnr{map}}(\SZ)$, but such points do not influence the integral in \eqref{Lemma.LB.2} and are neglected.} Hence, $\Cij(\SZ) \cap \mcR_{k}^{\tnr{map}}(\SZ) = \varnothing$ and \eqref{Lemma.LB.2} can be written as
\begin{align}\label{Lemma.LB.3}
\tpMAP{i}{j}(\SZ) \geq \int_{\mcH_{ij}^{\tnr{map}}(\SZ)\,\cap\,\Cij(\SZ)} \pdf(\by|\bx_{i})\, \tnr{d}\by.
\end{align}

For the third step, we return to \eqref{r.ineq}, which holds for any pair $i,j\in\mcIX$ with $j\neq i$. In the special case when $\deltaij = \MED$, \eqref{r.ineq} and \eqref{delta.ij} yield
\begin{align}\label{orthotope.ineq}
\frac{\MED}{2}-\frac{r(\SZ)}{\sqrt{N}} < \Delta_{ij}(\SZ) < \frac{\MED}{2}+\frac{r(\SZ)}{\sqrt{N}}.
\end{align}
Since $\Delta_{ij}(\SZ)$ gives the distance between $\bx_{i}$ and $\mcR_{j}^{\tnr{map}}(\SZ)$, and $\MED/2 \pm r(\SZ)/\sqrt{N}$ gives the distance between $\bx_{i}$ and two opposite facets of the hypercube $\Cij(\SZ)$, \eqref{orthotope.ineq} implies that $\mcH_{ij}^{\tnr{map}}(\SZ)\cap\Cij(\SZ)$ is a (nonempty) orthotope with thickness $\MED/2 + r(\SZ)/\sqrt{N}-\Delta_{ij}(\SZ)$. Carrying out the integration in \eqref{Lemma.LB.3} over this orthotope gives the second case of \eqref{Lemma.LB.eq}, which completes the proof of \lemmaref{Lemma.LB} for MAP detection.

The proof for ML detection is obtained similarly. The hypercube $\Cij(\SZ)$ is defined in the same way as before, and the analysis is identical up to \eqref{empty.eq.3}. Equation \eqref{empty.eq.4} is replaced by $\|\by-\bx_k\|^2-\|\by-\bx_i\|^2 \ge 0$, which via \eqref{pdf.channel} and \eqref{ml.j.region} implies $\by \notin \mcR_{k}^{\tnr{ml}}(\SZ)$. Hence, \eqref{Lemma.LB.3} is still valid and so is \eqref{orthotope.ineq}, except that $\Delta_{ij}(\SZ)$ in \eqref{orthotope.ineq} is now equal to $\MED/2$ in accordance with \eqref{delta.ij}. The thickness of $\mcH_{ij}^{\tnr{ml}}(\SZ)\cap\Cij(\SZ)$ is thus $r(\SZ)/\sqrt{N}$, which proves \eqref{Lemma.LB.eq} for ML detection.

\section{Proof of Theorem~\ref{EP.Asym.Theo}}\label{Appendix.EP.Asym.Theo}
To prove \theoref{EP.Asym.Theo}, we first use \eqref{P} to obtain
\begin{align}\label{proof.1}
\limsnszero{
\frac{P(\SZ)}{\QF\bigl(\frac{\MED}{2\SZ}\bigr)}
}
= \sumi \Pxi \sumjni h_{ij} 
\limsnszero{
\frac{\tp{i}{j}(\SZ)}{\QF\bigl(\frac{\MED}{2\SZ}\bigr)}
}.
\end{align}
As will become apparent later, the limit on the right hand side of \eqref{proof.1} exists and, hence, so does the limit on the left hand side. To calculate the limit in the right hand side of \eqref{proof.1}, we will sandwich it using  \lemmasref{Lemma.UB}{Lemma.LB}.

For MAP detection, we first study the asymptotic behavior of the upper bound in \lemmaref{Lemma.UB}
\begin{align}
\nonumber
&\limsnszero{\frac{\tpMAP{i}{j}(\SZ)}{\QF\bigl(\frac{\MED}{2\SZ}\bigr)}}\\
&\le 
\label{B.3}
\limsnszero{
\frac{\QF\bigl(\frac{\Delta_{ij}(\SZ)}{\SZ}\bigr)}{\QF\bigl(\frac{\MED}{2\SZ}\bigr)}
}\\
& =
\limsnszero{
\frac{
\QF\bigl(\frac{\deltaij}{2\SZ}+\frac{\SZ\log({\Pxi}/{\Pxj})}{\deltaij}\bigr)
}
{
\QF\bigl(\frac{\MED}{2\SZ}\bigr)}
}\\
\label{B.35}
& =
\nonumber
\limsnszero{
\frac{\frac{\deltaij}{2\SZ^{2}}-\frac{\log({\Pxi}/{\Pxj})}{\deltaij}}{\frac{\MED}{2\SZ^{2}}} }
\cdot
\\
&
\,\,
\mathrm{exp}\Biggl(
-\frac{\deltaij^{2}}{8\SZ^{2}}
-\frac{(\SZ\log({\Pxi}/{\Pxj}))^{2}}{2\deltaij^{2}}
-\frac{\log({\Pxi}/{\Pxj})}{2}
+\frac{\MED^{2}}{8\SZ^{2}}\Biggl)
\\
& = 
\limsnszero{
\frac{\deltaij}{\MED}\sqrt{\frac{p_j}{p_i}}
\explow{
-\frac{\deltaij^{2}-\MED^{2}}{8\SZ^{2}}}
}\\
\label{B.4}
&=
\begin{cases}
0, & \textnormal{if $\deltaij>\MED$},\\
\sqrt{\frac{\Pxj}{\Pxi}},& \textnormal{if $\deltaij=\MED$},
\end{cases}
\end{align}
where \eqref{B.35} follows from l'H\^{o}pital's rule {\cite[Sec.~11.2]{Marsden85}}.

Next, we study the asymptotic behavior of the lower bound in \lemmaref{Lemma.LB} for $\deltaij=\MED$ (the lower bound \eqref{Lemma.LB.eq} is zero for $\deltaij>\MED$). Assuming that all the limits exist, we obtain
\begin{align}
\nonumber
&\limsnszero{\frac{\tpMAP{i}{j}(\SZ)}{\QF\bigl(\frac{\MED}{2\SZ}\bigr)}}\\
& \hspace{-2ex} \ge
\limsnszero{
\frac{
\left(\QF\Bigl(\frac{\Delta_{ij}(\SZ)}{\SZ}\Bigr)-\QF\Bigl(\frac{\MED}{2\SZ}+\frac{r(\SZ)}{\sqrt{N}\SZ}\Bigr)\right)\left(1-2\QF\left(\frac{r(\SZ)}{\sqrt{N}\SZ}\right)\right)^{N-1}
}
{
\QF\bigl(\frac{\MED}{2\SZ}\bigr)
}
}\\
\label{B.5}
& \hspace{-2ex}=
\nonumber
\left[
\limsnszero{\frac{\QF\Bigl(\frac{\Delta_{ij}(\SZ)}{\SZ}\Bigr)}{\QF\bigl(\frac{\MED}{2\SZ}\bigr)}}
-
\limsnszero{\frac{\QF\Bigl(\frac{\MED+2r(\SZ)/\sqrt{N}}{2\SZ}\Bigr)}{\QF\bigl(\frac{\MED}{2\SZ}\bigr)}}
\right]\cdot\\
&
\qquad\qquad\qquad\qquad\quad\,\,
\limsnszero{\left(1-2\QF\left(\frac{r(\SZ)}{\sqrt{N}\SZ}\right)\right)^{N-1}}.
\end{align}
The first limit in \eqref{B.5} is the same as in \eqref{B.3}--\eqref{B.4}. The second limit is zero and the last limit is one because by \eqref{r}, $\limsnszero{r(\SZ)} = \MED/(2(1+\sqrt{3}))$. Hence, all limits exist and asymptotically, both lower and upper bounds converge to \eqref{B.4}. Using this in \eqref{proof.1} gives
\begin{align}\label{proof.2}
\limsnszero{
\frac{P(\SZ)}{\QF\bigl(\frac{\MED}{2\SZ}\bigr)}
}
= \sumi \Pxi \sumjnid h_{ij} 
\sqrt{\frac{\Pxj}{\Pxi}},
\end{align}
which completes the proof for MAP detection.

The proof for ML detection follows similar steps. Substituting $\Delta_{ij}(\SZ)=\deltaij/2$ from \eqref{delta.ij} into \eqref{B.3} yields
\begin{align}
\limsnszero{\frac{\tpML{i}{j}(\SZ)}{\QF\bigl(\frac{\MED}{2\SZ}\bigr)}}
&\le
\limsnszero{
\frac{\QF\bigl(\frac{\deltaij}{2\SZ}\bigr)}{\QF\bigl(\frac{\MED}{2\SZ}\bigr)}
}\\
&=
\begin{cases}
0, & \textnormal{if $\deltaij>\MED$},\\
1, & \textnormal{if $\deltaij=\MED$}.
\label{B.3.ML}
\end{cases}
\end{align}
The asymptotic expression for the lower bound in \eqref{B.5} holds unchanged in the ML case too. In this case, the first limit is given by \eqref{B.3.ML}, the second is zero, and the third is one. This combined with \eqref{proof.1} completes the proof for ML detection.

\section{Proof of Corollary~\ref{EP.MAP_vs_ML.Theo}}\label{Appendix.EP.MAP_vs_ML.Theo}
Equations \eqref{EP.MAP_vs_ML}--\eqref{R} follow immediately from \eqref{EP.Asym}--\eqref{EP.B}.
To prove $R\leq 1$, we need to prove 
\begin{align}
\sumijd h_{ij}\Pxi-\sumijd h_{ij}\sqrt{\Pxj\Pxi}\geq 0. 
\end{align}
Using
 $h_{ij}=h_{ji}$ and $\delta_{ij}=\delta_{ji}$, we obtain
\begin{align}\label{B.C.inequality}
\sumijd h_{ij}(\Pxi - \sqrt{\Pxj\Pxi})
\nonumber
			& = \frac{1}{2} \sumijd h_{ij}(\Pxi-\sqrt{\Pxj\Pxi})+\\
			& \quad\,\, \frac{1}{2} \sumjid h_{ji}(\Pxj-\sqrt{\Pxi\Pxj})\\
			& = \frac{1}{2} \sumijd h_{ij}(\Pxi+ \Pxj - 2\sqrt{\Pxj\Pxi})\\
			& = \frac{1}{2} \sumijd h_{ij}(\sqrt{\Pxi}-\sqrt{\Pxj})^{2}\\
			& \geq 0,
\end{align}
which holds with equality if and only if $\Pxi=\Pxj, \, \forall i,j:\deltaij=\MED$.

\balance
\bibliography{IEEEabrv,references_MAP}

\begin{thebibliography}{10}
\providecommand{\url}[1]{#1}
\csname url@samestyle\endcsname
\providecommand{\newblock}{\relax}
\providecommand{\bibinfo}[2]{#2}
\providecommand{\BIBentrySTDinterwordspacing}{\spaceskip=0pt\relax}
\providecommand{\BIBentryALTinterwordstretchfactor}{4}
\providecommand{\BIBentryALTinterwordspacing}{\spaceskip=\fontdimen2\font plus
\BIBentryALTinterwordstretchfactor\fontdimen3\font minus
  \fontdimen4\font\relax}
\providecommand{\BIBforeignlanguage}[2]{{%
\expandafter\ifx\csname l@#1\endcsname\relax
\typeout{** WARNING: IEEEtran.bst: No hyphenation pattern has been}%
\typeout{** loaded for the language `#1'. Using the pattern for}%
\typeout{** the default language instead.}%
\else
\language=\csname l@#1\endcsname
\fi
#2}}
\providecommand{\BIBdecl}{\relax}
\BIBdecl

\bibitem{Gilbert52}
E.~N. Gilbert, ``A comparison of signalling alphabets,'' \emph{Bell System
  Technical Journal}, vol.~31, no.~3, pp. 504--522, May 1952.

\bibitem{Wozencraft65_Book}
J.~M. Wozencraft and I.~M. Jacobs, \emph{Principles of Communication
  Engineering}.\hskip 1em plus 0.5em minus 0.4em\relax John Wiley \& Sons,
  1965.

\bibitem{Jacobs67}
I.~Jacobs, ``Comparison of $m$-ary modulation systems,'' \emph{Bell System
  Technical Journal}, vol.~46, no.~5, pp. 843--864, May-June 1967.

\bibitem{Foschini74}
G.~J. Foschini, R.~D. Gitlin, and S.~B. Weinstein, ``Optimization of
  two-dimensional signal constellations in the presence of {G}aussian noise,''
  \emph{IEEE Trans. Commun.}, vol.~22, no.~1, pp. 28--38, Jan. 1974.

\bibitem{Smith75}
J.~G. Smith, ``Odd-bit quadrature amplitude-shift keying,'' \emph{IEEE Trans.
  Commun.}, vol.~23, pp. 385--389, Mar. 1975.

\bibitem{Forney89a}
G.~D. {Forney, Jr.} and L.-F. Wei, ``Multidimensional constellations---{Part
  I}: Introduction, figures of merit, and generalized cross constellations,''
  \emph{IEEE J. Sel. Areas Commun.}, vol.~7, no.~6, pp. 877--892, Aug. 1989.

\bibitem{Kschischang93}
F.~R. Kschischang and S.~Pasupathy, ``Optimal nonuniform signaling for
  {G}aussian channels,'' \emph{IEEE Trans. Inf. Theory}, vol.~39, no.~39, pp.
  913--929, May 1993.

\bibitem{Ivanov13a}
M.~Ivanov, F.~Br\"{a}nnstr\"{o}m, A.~Alvarado, and E.~Agrell, ``On the exact
  {BER} of bit-wise demodulators for one-dimensional constellations,''
  \emph{IEEE Trans. Commun.}, vol.~61, no.~4, pp. 1450--1459, Apr. 2013.

\bibitem{Valenti12}
M.~Valenti and X.~Xiang, ``Constellation shaping for bit-interleaved {LDPC}
  coded {APSK},'' \emph{IEEE Trans. Commun.}, vol.~60, no.~10, pp. 2960--2970,
  Oct. 2012.

\bibitem{Lassing03}
J.~Lassing, E.~G. Str\"{o}m, E.~Agrell, and T.~Ottosson, ``Computation of the
  exact bit error rate of coherent {M}-ary {PSK} with {Gray} code bit
  mapping,'' \emph{IEEE Trans. Commun.}, vol.~51, no.~11, pp. 1758--1760, Nov.
  2003.

\bibitem{Agrell04}
E.~Agrell, J.~Lassing, E.~G. Str\"{o}m, and T.~Ottosson, ``On the optimality of
  the binary reflected {G}ray code,'' \emph{IEEE Trans. Inf. Theory}, vol.~50,
  no.~12, pp. 3170--3182, Dec. 2004.

\bibitem{Alvarado12b}
A.~Alvarado, F.~Br\"{a}nnstr\"{o}m, E.~Agrell, and T.~Koch, ``High-{SNR}
  asymptotics of mutual information for discrete constellations with
  applications to {BICM},'' \emph{IEEE Trans. Inf. Theory}, vol.~60, no.~2, pp.
  1061--1076, Feb. 2014.

\bibitem{Zetterberg77}
L.~H. Zetterberg and H.~Br\"andstr\"om, ``Codes for combined phase and
  amplitude modulated signals in a four-dimensional space,'' \emph{IEEE Trans.
  Commun.}, vol. COM-25, no.~9, pp. 943--950, Sept. 1977.

\bibitem{Steiner94}
M.~Steiner, ``The strong simplex conjecture is false,'' \emph{IEEE Trans. Inf.
  Theory}, vol.~40, no.~3, pp. 721--731, May 1994.

\bibitem{Thomas74}
C.~M. Thomas, M.~Y. Weidner, and S.~H. Durrani, ``Digital amplitude-phase
  keying with {$M$-ary} alphabets,'' \emph{IEEE Trans. Commun.}, vol. COM-22,
  no.~2, pp. 168--180, Feb. 1974.

\bibitem{Marsden85}
J.~Marsden and A.~Weinstein, \emph{Calculus II}, 2nd~ed.\hskip 1em plus 0.5em
  minus 0.4em\relax New York, NY: Springer, 1985.

\end{thebibliography}
\bibliographystyle{IEEEtran}

\begin{IEEEbiographynophoto}{Alex Alvarado}
(S'06--M'11--SM'15) was born in Quell\'on, on the island of Chilo\'e, Chile. He received his Electronics Engineer degree (Ingeniero Civil Electr\'onico) and his M.Sc. degree (Mag\'ister en Ciencias de la Ingenier\'ia Electr\'onica) from Universidad T\'ecnica Federico Santa Mar\'ia, Valpara\'iso, Chile, in 2003 and 2005, respectively. He obtained the degree of Licentiate of Engineering (Teknologie Licentiatexamen) in 2008 and his PhD degree in 2011, both of them from Chalmers University of Technology, Gothenburg, Sweden.

Dr. Alvarado is an Assistant Professor at the Signal Processing Systems (SPS) Group, Department of Electrical Engineering, Eindhoven University of Technology (TU/e), The Netherlands. During 2014–-2016, he was a Senior Research Associate at the Optical Networks Group, University College London, United Kingdom. In 2012--2014 he was a Marie Curie Intra-European Fellow at the University of Cambridge, United Kingdom, and during 2011--2012 he was a Newton International Fellow at the same institution. Dr. Alvarado is a recipient of the 2009 IEEE Information Theory Workshop Best Poster Award, the 2013 IEEE Communication Theory Workshop Best Poster Award, and the 2015 IEEE Transaction on Communications Exemplary Reviewer Award. He is a senior member of the IEEE and an associate editor for IEEE Transactions on Communications (Optical Coded Modulation and Information Theory). His general research interests are in the areas of digital communications, coding, and information theory.
\end{IEEEbiographynophoto}

\begin{IEEEbiographynophoto}{Erik Agrell} (M'99--SM'02) received the Ph.D.~degree in information theory in 1997 from Chalmers University of Technology, Sweden.

From 1997 to 1999, he was a Postdoctoral Researcher with the University of California, San Diego and the University of Illinois at Urbana-Champaign. In 1999, he joined the faculty of Chalmers University of Technology, where he is a Professor in Communication Systems since 2009. In 2010, he cofounded the Fiber-Optic Communications Research Center (FORCE) at Chalmers, where he leads the signals and systems research area. He was a Visiting Professor at University College London in 2014--2017. His research interests belong to the fields of information theory, coding theory, and digital communications, and his favorite applications are found in optical communications.

Prof.~Agrell served as Publications Editor for the IEEE Transactions on Information Theory from 1999 to 2002 and as Associate Editor for the IEEE Transactions on Communications from 2012 to 2015. He is a recipient of the 1990 John Ericsson Medal, the 2009 ITW Best Poster Award, the 2011 GlobeCom Best Paper Award, the 2013 CTW Best Poster Award, and the 2013 Chalmers Supervisor of the Year Award.
\end{IEEEbiographynophoto}

\begin{IEEEbiographynophoto}{Fredrik Br\"annstr\"om} (S'98--M'05) received the M.Sc.~degree in electrical engineering from Lule\aa~University of Technology, Lule\aa, Sweden, in 1998, and the Ph.D.~degree in communication theory from the Department of Computer Engineering, Chalmers University of Technology, Gothenburg, Sweden, in 2004. 

From 2004 to 2006, he was a Post-Doctoral Researcher with the Communication Systems Group, Department of Signals and Systems, Chalmers University of Technology. From 2006 to 2010, he was a Principal Design Engineer with Quantenna Communications, Inc., Fremont, CA, USA. In 2010, he returned to Chalmers University of Technology, where he is currently a Professor with the Communication Systems Group at the Department of Electrical Engineering. His research interests in communication and information theory include code design, coded modulation, labelings, and coding for distributed storage, as well as algorithms, resource allocation, synchronization, and protocol design for vehicular communication systems. He was a recipient of the 2013 IEEE Communication Theory Workshop Best Poster Award. In 2014, he received the Department of Signals and Systems Best Teacher Award. He has co-authored the papers that received the 2016 and 2017 IEEE Sweden VT-COM-IT Joint Chapter Best Student Conference Paper Award.
\end{IEEEbiographynophoto}

\end{document}